\documentclass[11pt,twoside,onecolumn]{article}
\usepackage[]{latexsym}
\usepackage[dvips]{epsfig}
\usepackage{amsmath,amssymb,amsfonts,mathrsfs}
\title{\bf Einstein static universe as a brane in extra dimensions}
\author{A. Gruppuso$^{a,b,c}$\thanks{gruppuso@bo.iasf.cnr.it} ,$\ $
E. Roessl$^{a}$\thanks{ewald.roessl@epfl.ch} $\ $
and M. Shaposhnikov$^{a}$\thanks{mikhail.shaposhnikov@epfl.ch}
\\
\\
{\em $^{a}$ Institut de th\'eorie des ph\'enom\`enes physiques (ITP)} \\
{\em Laboratoire de Physique des Particules et Cosmologie (LPPC)}\\
{\em \'Ecole polytechnique f\'ed\'erale de Lausanne} \\
{\em CH-1015 Lausanne, Switzerland}
\\
\\
{\em $^{b}$ IASF/CNR, Istituto di Astrofisica Spaziale e Fisica Cosmica} \\
{\em Sezione di Bologna} \\
{\em Consiglio Nazionale delle Ricerche} \\
{\em via Gobetti 101, I-40129 Bologna - Italy}\\
and
\\
{\em $^{c}$ Dipartimento di Fisica, Universit\`a di Bologna and
I.N.F.N., Sezione di Bologna,} \\
{\em via Irnerio 46, 40126, Bologna, Italy}}

\begin{document}

\maketitle
\thispagestyle{empty}
\begin{abstract}
We present a brane-world scenario in which two regions of $AdS_5$ space-time are glued together 
along a 3-brane with constant positive curvature such that {\em all} 
spatial dimensions form a compact manifold of topology $S^4$. It turns out 
that the induced geometry on the brane is given by Einstein's static universe.
It is possible to achieve an anisotropy of the manifold which allows for 
a huge hierarchy between the size of the extra dimension $R$ and the size of the observable universe 
$R_U$ at present. This anisotropy is also at the origin of a very peculiar property of our model: 
the physical distance between {\em any two points} on the brane is of the order of the size of the extra 
dimension $R$ regardless of their distance measured with the use of the induced metric on the brane.
In an intermediate distance regime $R \ll r \ll R_U$ gravity on the brane 
is shown to be effectively $4$-dimensional, 
with corresponding large distance corrections, in complete analogy with the Randall-Sundrum II model. 
For very large distances $r \sim R_U$ we recover gravity in Einstein's static universe. However,
in contrast to the Randall-Sundrum II model the difference in topology has the advantage of giving rise to a 
geodesically complete space. 
\end{abstract}

\newpage
\raggedbottom
\setcounter{page}{1}

\section{Introduction}
\setcounter{equation}{0}
\label{intro}

Recent suggestions that large \cite{Antoniadis:1990ew}--\cite{Antoniadis:1998ig} or infinite 
\cite{Rubakov:bb}--\cite{Randall:1999vf} extra dimensions are not necessarily in conflict with present
observations provide new opportunities for addressing several outstanding problems of modern theoretical 
physics like the hierarchy problem \cite{Antoniadis:1990ew}--\cite{Antoniadis:1998ig},\cite{Randall:1999ee}, 
\cite{Cohen:1999ia} or the cosmological constant problem \cite{Rubakov:1983bz}--\cite{Randjbar-Daemi:1985wg}.
In the course of this development and inspired by string theoretical arguments \cite{Polchinski:1995mt}, 
the notion of brane world scenarios emerged in which the usual Standard Model fields are supposed to 
be confined to a so-called $3$-brane, a $4$-dimensional sub-manifold of some higher-dimensional space-time.
As shown in \cite{Randall:1999vf} also gravity can appear to be effectively $4$-dimensional for a 
brane-bound observer provided the conventional scheme of Kaluza-Klein compactification 
\cite{Kaluza:tu}--\cite{KK} is replaced by a compactification using non-factorizable (also called warped)
geometries (see also \cite{Rubakov:1983bz}). 
This triggered an immense research activity in theories involving $3$-branes with
interests ranging from elucidating the global space-time structure of brane world scenarios, properties 
of gravity, cosmology and brane cosmological perturbations, generalizations to higher dimensions etc.
While the possibilities are rich, realistic scenarios remain rare. For example the simple Randall-Sundrum-II
model \cite{Randall:1999vf} faces the problem of being geodesically incomplete 
\cite{Rubakov:2001kp,Muck:2000bb,Gregory:2000rh}. 
It is therefore reasonable to look for adequate alternatives or generalizations 
to the Randall-Sundrum-II model which avoid the 
above mentioned problems while sharing its pleasant feature of the effective $4$-dimensional low
energy gravity on the brane.

In this paper we present a $5$-dimensional brane-world model which solves the geodesic
incompleteness of the Randall-Sundrum II model while preserving its phenomenological 
properties concerning the localization of gravity. 

We try to illustrate our motivation for considering a particular geometry by using the simple 
picture of a domain structure in extra dimensions resulting from a spontaneously broken 
discrete symmetry. The associated Higgs-field takes different values in regions separated by a domain wall, 
which restricts the possibilities of combining domain walls depending on the global topology of the 
space-time under consideration. We concentrate on the $5$-dimensional case, see 
\cite{Cohen:1999ia},\cite{Chodos:1999zt}--\cite{Gherghetta:2000jf} for higher dimensional constructions.

\begin{figure}[htpb]
\centerline{\epsfxsize=3.0in\epsfbox{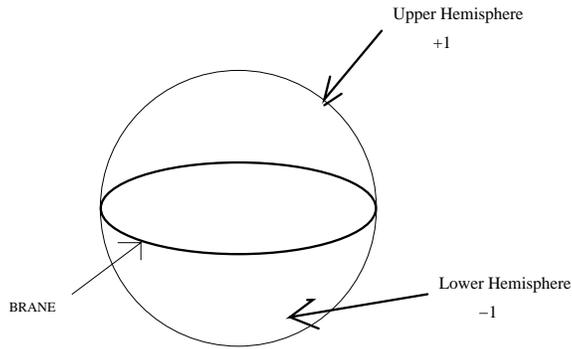}}
\caption{Compact topology. $S^3$-brane embedded in $S^4$. $+1$ and $-1$ symbolically 
represent the different vacuum expectation values taken by the Higgs field in different regions of space.}
\label{ct}
\end{figure}

Let us analyze a few simple cases:
\begin{itemize}
{\item Non-compact extra dimension}: $y \in (-\infty, + \infty)$.
In this case, by choosing the origin $y=0$ to coincide with the position of the brane, we 
obtain two non-overlapping regions
$(-\infty, 0) $ and $(0, + \infty)$ and we can thus consistently have one brane in such a theory. 
An example for this configuration is provided by the Randall-Sundrum II model \cite{Randall:1999vf}.
{\item Compact extra dimension}: $y \in [0 , 2 \pi]$. We now consider two possibilities, depending on
the spatial topology of our manifold:
\begin{itemize}
{\item{\bf (a)}} $\mathbb{R}^3 \times S^1$: If the ordinary dimensions are supposed to be 
non-compact, it is not possible
to consistently put only one brane in the extra dimensions. At least two branes are needed.
{\item{\bf (b)}} $S^4$: If {\it all} spatial coordinates are part of a compact
manifold, the simple picture shown in Fig.~\ref{ct} seems to suggest that it is possible to consistently put 
a single brane in the bulk space-time.
\end{itemize}
\end{itemize}
To the best of our knowledge case {\bf (b)} has not yet been considered in the literature and 
our aim is to present such a construction.

This paper is organized as follows: in Section \ref{one} we discuss the basic geometric and 
topological properties of our brane-world scenario like  Einstein equations, junction
conditions and the distance hierarchy between the extra dimension $R$ and the observable universe 
$R_U$. Section \ref{two} is dedicated to the study of geodesics and the demonstration that our model 
does not suffer from being geodesically incomplete. In section \ref{three} we present a detailed computation 
of the propagator of a massless scalar field in the given background serving as an easy, phenomenological 
approach to the study of gravity. We discuss the behavior of the two-point function in three different 
distance regimes. It turns out that the computations are rather technical and we therefore collect large 
parts of it in four appendices. We eventually draw conclusions in section \ref{four}.

\section{The background equations}
\setcounter{equation}{0}
\label{one}
The aim of this section is to present the topology and geometry of the brane-world model 
which we set out to study but also to motivate how this particular model emerged through 
imposing several physical conditions on the more general set of solutions.

\subsection{Einstein equations}

We would like to construct a space-time with all spatial dimensions being part of the 
same compact manifold. One more motivation for this lies in a possible solution of the strong CP
problem within theories with extra dimensions \cite{Khlebnikov:1987zg}.  
The overall topology would then be given by $\mathbb{R}\times \mathbb{K}$,
where $\mathbb{R}$ represents the time-coordinate and $\mathbb{K}$ any compact manifold. 
As announced, we will restrict ourselves to the case of one extra dimension and an induced metric 
on the brane characterized by a spatial component of geometry $S^3$. The idea is to combine two 
$5$-dimensional regions of space-time dominated by a cosmological constant $\Lambda$ in such a 
way that the border of the two regions can be identified with a $3$-brane, constituting 
our observable universe. As pointed out in \cite{Khlebnikov:1987zg} the manifold $\mathbb{K}$ has to 
be highly 
anisotropic in order to single out the small extra dimension from the three usual ones. It
is not a priory clear whether the Einstein equations allow for such solutions at all and if so whether 
gravity can be localized on the brane in such a setup.

We choose the following ansatz for the $5$-dimensional metric consistent with the above 
requirements:\footnote{Our conventions for the metric and the Riemann tensor are those of reference \cite{wheeler}.}
\begin{eqnarray} \label{metric}
  ds^2 = g_{M N} dx^M dx^N=-\sigma^2(\theta)\, dt^2 + R_U^2 \, \gamma ^2(\theta) 
         \, d\Omega^2_3 + R^2 \, d\theta^2 \, ,
\end{eqnarray}
where $\theta \in [-\frac{\pi}{2},\frac{\pi}{2}]$ denotes the extra coordinate and 
$R$ and $R_U$ are constants representing the size 
of the extra dimension and the size of the observable universe at present, respectively.
$d\Omega^2_3$ denotes the line element of a $3$-sphere:
\begin{equation} \label{S3metric}
  d\Omega^2_3 =d\varphi^2_1 + \sin ^2 \varphi_1 \, d\varphi^2_2 + \sin ^2 \varphi_1\, \sin ^2 \varphi_2\,  
  d\varphi^2_3 \, ,
\end{equation}
where $\varphi_1, \varphi_2$ belong to the interval $\left[ 0, \pi \right]$, $ \varphi_3 $
to $\left[ 0, 2 \pi \right]$.
Capital latin letters $M,N,..$ will range from $0$ to $4$.
In order to obtain a compact space, we require $\gamma(\pm\frac{\pi}{2})=0$.

Metrics which can locally be put into the form (\ref{metric}) are sometimes referred to as 
{\em asymmetrically warped} metrics due to two different functions ($\sigma$ and $\gamma$ in our 
notations) multiplying the temporal and spatial part of the $4$-dimensional coordinate 
differentials. Their relevance in connection with $4$-dimensional Lorentz-violation at 
high energies was first pointed out in \cite{Visser:1985qm}. More recent discussions of this subject 
can be found in \cite{Csaki:2000dm}--\cite{Dubovsky:2001fj} or \cite{Rubakov:2001kp} and 
references therein. A common prediction of theories of this kind is that dispersion 
relations get modified at high energies. 
For a field-theoretical discussion of Lorentz violating effects 
in the context of the Standard Model of particle physics see~\cite{Coleman:1998ti}.
  
The Einstein equations in 5 dimensions with a bulk cosmological constant 
$\Lambda$ and a stress-energy tensor $T_{M N}$ take the following form:
\begin{eqnarray} \label{einsteineq}
  R_{M N} -\frac{1}{2} \, R \, g_{M N} + \Lambda \, g_{M N}
  = \frac{8 \pi}{M^3} \, T_{M N} \, ,
\end{eqnarray}
where $M$ is the fundamental scale of gravity.
We choose to parametrize the stress-energy tensor in the following way, consistent with the symmetries
of the metric:
\begin{eqnarray} \label{stress}
  T^0_{\;\;0}=\epsilon_0(\theta) < 0\, , \quad
  T^i_{\;\;j}= \delta^i_j\,\epsilon(\theta)\, , \quad
  T^\theta_{\;\;\theta}=\epsilon_\theta(\theta) \, ,
\end{eqnarray}
where as indicated the above diagonal components depend only on the extra coordinate $\theta$.
Lower case latin indices $i,j$ label the coordinates on $S^3$. 

Using the metric ansatz (\ref{metric}) together with the stress-energy tensor (\ref{stress}) 
the Einstein equations (\ref{einsteineq}) become
\begin{eqnarray}
\frac{3}{R^2} \left[ \left( \frac{\gamma^{\prime}}{\gamma}\right)^2 + \frac{\gamma^{\prime \prime}}{\gamma} -
 \left( \frac{R}{R_U} \right)^2 \frac{1}{\gamma ^2} \right] + \Lambda &=& \frac{8 \pi}{M^3} 
  \epsilon_0 \, , \label{00} \\
  \frac{1}{R^2} \left[ \left( \frac{\gamma ^{\prime}}{\gamma}\right)^2 +
  2 \frac{\gamma^{\prime}}{\gamma} \frac{\sigma^{\prime}}{\sigma} 
  +2 \frac{\gamma ^{\prime \prime}}{\gamma} + \frac{\sigma^{\prime \prime}}{\sigma} -
  \left( \frac{R}{R_U} \right)^2 \frac{1}{\gamma^2} \right] + \Lambda &=& \frac{8 \pi}{M^3} 
  \epsilon \, , \label{ii} \\
  \frac{3}{R^2} \left[ \left( \frac{\gamma^{\prime}}{\gamma}\right)^2 +
  \frac{\gamma^{\prime}}{\gamma} \frac{\sigma^{\prime}}{\sigma} -
  \left( \frac{R}{R_U} \right)^2 \frac{1}{\gamma ^2} \right] + \Lambda &=& \frac{8 \pi}{ M^3} 
  \epsilon_{\theta} \, , \label{thetatheta}
\end{eqnarray}
where $^{\prime}$ denotes differentiation with respect to $\theta$.
The conservation of stress-energy, or equivalently, the Bianchi-identities lead to the following constraint 
relating the three independent components $\epsilon_0$, $\epsilon$ and $\epsilon_\theta$:
\begin{eqnarray}\label{constraint}
  \epsilon^\prime_{\theta} + \left( \frac{\sigma^{\prime}}{\sigma} +
  3 \frac{\gamma^{\prime}}{\gamma}\right) \, \epsilon_{\theta} -
  \frac{\sigma^{\prime}}{\sigma}\, \epsilon_{0} - 3 \, \frac{\gamma^{\prime}}{\gamma} \epsilon =0 \, .
\end{eqnarray}

\subsection{Vacuum solution}
In this paper we do not intend to provide a field theoretical model which could 
generate the geometry we are about to describe. Our aim is to study a singular brane, located at
$\theta_b=0$ separating two vacuum regions in the bulk. Of course the solutions to the Einstein 
equations in vacuum with a cosmological constant term are nothing but the familiar 
de-Sitter ($\Lambda>0$), Minkowski ($\Lambda=0$) and anti-de-Sitter ($\Lambda<0$) space-times.
To see how these geometries can be recovered using the line element (\ref{metric}) we set
the right hand sides of (\ref{00})-(\ref{thetatheta}) equal to zero and subtract eq.~(\ref{thetatheta}) 
from eq.~(\ref{00}) to obtain $\sigma = c \, \gamma^{\prime}$, where $c$ is a constant.
Putting this back in eqs.~(\ref{00})-(\ref{thetatheta}), we are left with only one independent differential
equation  
\begin{align}\label{00n}
  \frac{3}{R^2} \left[ \left( \frac{\gamma^{\prime}}{\gamma}\right)^2 + 
  \frac{\gamma^{\prime \prime}}{\gamma} -
  \left( \frac{R}{R_U} \right)^2 \frac{1}{\gamma^2} \right] + \Lambda = 0.
\end{align}
The solution of (\ref{00n}) is trivial and by means of $\sigma = c \, \gamma^{\prime}$ we find 
\begin{equation} \label{sigmaandgamma}
  \gamma(\theta) =
  \frac{\sinh \left[\omega\left(\frac{\pi}{2}-|\theta| \right)\right]}
     {\sinh (\omega \frac{\pi}{2})}\, ,\qquad
  \sigma (\theta) =
  \frac{\cosh \left[\omega\left(\frac{\pi}{2}-|\theta| \right)\right]}
    {\cosh (\omega \frac{\pi}{2})}\, ,
\end{equation}
provided that $R_U = R \sinh (\omega \frac{\pi}{2})/\omega$ and $\omega^2 =  -\Lambda  R^2 /6 $.
Notice that the solution (\ref{sigmaandgamma}) does not contain any integration constant because we 
already imposed the boundary conditions $\gamma(\pm\frac{\pi}{2})=0$  
and $\gamma(0)=\sigma(0)=1$. Moreover, we chose the whole setup to be symmetric under the transformation
$\theta \to -\theta$.
From (\ref{sigmaandgamma}) it is now obvious that locally (for $\theta>0$ and $\theta<0$) 
the line element (\ref{metric}) correctly describes de-Sitter, Minkowski and anti-de-Sitter space-times for 
imaginary, zero and real values of $\omega$, respectively. For our purposes only the $AdS$ solution will be
of any interest as we will see in the following section. A simple change of coordinates starting from
(\ref{metric}) and (\ref{sigmaandgamma}) shows that the $AdS$-radius in our notations is given by 
$R_{AdS}=R/\omega=\sqrt{-6/\Lambda}$.\footnote{See also footnote \ref{changecoords}.} 
At this stage, the validity of the vacuum solution 
(\ref{sigmaandgamma}) is restricted to the bulk, since it is not even differentiable in the classical 
sense at $\theta=0$. In order to give sense to (\ref{sigmaandgamma}) for all values of $\theta$ we will 
have to allow for some singular distribution of stress-energy at $\theta=0$ and solve 
(\ref{00})-(\ref{thetatheta}) in the sense of distributions.

\subsection{The complete solution for a singular brane}
\label{SingularBrane}
As announced, we now refine our ansatz for the energy momentum tensor (\ref{stress}) 
to allow for a solution of eqs. (\ref{00})-(\ref{thetatheta}) in the whole interval 
$-\pi/2\leq\theta\leq\pi/2$:
\begin{align} \label{stressrefined}
  \epsilon_0 (\theta ) = c_0 \, \frac{\delta(\theta)}{R} \, ,\quad  
  \epsilon (\theta ) = c  \, \frac{\delta(\theta)}{R} \, ,\quad 
  \epsilon_{\theta} (\theta ) = c_\theta \, \frac{\delta(\theta)}{R} \, .
\end{align}
After replacing the above components (\ref{stressrefined}) in eqs.~(\ref{00})-(\ref{thetatheta}) and 
integrating over $\theta$ from $-\eta$ to $\eta$, followed by the limit $\eta \to 0$ we find:
\begin{align}  
  :\frac{\gamma^\prime}{\gamma}: \;\; &= \frac{8 \pi R}{3 M^3} c_0 \label{FineTuneI} \, ,\\
  :\frac{\sigma^\prime}{\sigma}: \;+ \;2:\frac{\gamma^\prime}{\gamma}: \;\; &=\frac{8\pi R}{M^3}c 
  \label{FineTuneII} \, ,\\
  0&=c_\theta \, ,
\end{align}
where the symbol $: \ldots :$ is used to denote the jump of a quantity across the brane defined by:
\begin{equation} \label{::def}
  : f : \;\; \equiv \lim_{\eta\to0} \left[f(\eta)-f(-\eta)\right].
\end{equation}
Note that in the above step we made use of the identity
\begin{equation}  
  \frac{\gamma^{\prime \prime}}{\gamma} = \left(\frac{\gamma^\prime}{\gamma} \right)^\prime + 
  \left( \frac{\gamma^\prime}{\gamma} \right)^2 
\end{equation}
together with the continuity of $\gamma$ on the brane. Specifying (\ref{FineTuneI}) and (\ref{FineTuneII}) 
to (\ref{sigmaandgamma}) we have:
\begin{align}
  c_0 &= - \frac{3}{4\pi} M^3 \frac{\omega}{R} \, \coth \left(\omega \frac{\pi}{2}\right) \, ,\label{solc0}\\
  c_{\phantom{0}} &= - \frac{1}{4\pi} M^3 \frac{\omega}{R} \, 
   \left[ \tanh \left(\omega \frac{\pi}{2}\right)+2 \coth \left(\omega \frac{\pi}{2}\right)\right] 
   \, , \label{solc} \\
  c_\theta &= 0 \, . \label{solctheta}
\end{align}
Eqs.~(\ref{solc0}) and (\ref{solc}) relate the energy-density and the pressure of the singular brane to the bulk cosmological constant $\Lambda$ (via $\omega$), the size of the extra dimension $R$ 
and the fundamental scale of gravity $M$. With the above relations (\ref{solc0})-(\ref{solctheta}) we 
can now interpret (\ref{sigmaandgamma}) as a solution to the Einstein equations 
(\ref{00})-(\ref{thetatheta}) in the sense 
of distributions. Also the stress-energy conservation constraint 
(\ref{constraint}) is satisfied based on the identity $\delta(x) \, \mbox{sign}(x)=0$ again in the 
distributional sense. 

Note that $c_0/c \to 1$ in the limit $\omega \to \infty$. 
Moreover, for larger and larger values of $\omega$, $c_0$ and $c$ 
approach the brane tension of the Randall-Sundrum II model and the above eqs.~ (\ref{solc0}) and 
(\ref{solc}) merge to the equivalent relation in the Randall-Sundrum II case. This is no surprise since
taking the limit $\omega \to \infty$ corresponds to inflating and flattening the 3-brane 
so that we expect to recover the case of the flat Randall-Sundrum II brane.

We finish this section by the discussion of some physical properties of our manifold. We first  
observe that its spatial part is homeomorphic to a $4$-sphere $S^4$. 
This is obvious from the metric (\ref{metric}) and the explicit expression for $\gamma$ given in 
(\ref{sigmaandgamma}). Geometrically, however, our manifold differs from $S^4$ due to the high
anisotropy related to the smallness of the extra dimension.
The ratio of typical distance scales in the bulk and on the brane is given by
\begin{eqnarray} \label{hierarchy}
  \frac{R}{R_{U}}= \frac{\omega}{\sinh \left( \frac{\omega\pi}{2}\right)} \, .
\end{eqnarray}
It is now clear that the above ratio (\ref{hierarchy}) can only be made very small in the case of 
real $\omega$ ($AdS$-space-time). For the size of the observable universe we take the lower bound 
$R_U > 4 \, \mbox{Gpc}\sim 10^{28}\,\mbox{cm}$ while the size of the extra dimension is 
limited from above \cite{Hoyle:2000cv}: $R<10^{-2}\,\mbox{cm}$, leaving us with $\omega > 50$.

Finally we would like to point out a very peculiar property of the manifold under consideration: 
as it can immediately be deduced from the line element (\ref{metric}) and (\ref{sigmaandgamma}),
any two points on the brane are separated by not more than a distance of the order of $R$ regardless of their 
distance as measured by an observer on the brane using the induced metric.

\section{Geodesics}
\setcounter{equation}{0}
\label{two}

It is well known that the Randall Sundrum-II model is timelike and lightlike geodesically incomplete 
\cite{Rubakov:2001kp,Muck:2000bb,Gregory:2000rh} which means that there exists inextendible timelike and lightlike geodesics.\footnote{For a precise
definition of a geodesically incomplete space see e.g. \cite{Wald}.} An inextendible geodesic is a geodesic
parametrized by an affine parameter $\tau$ such that by using up only a finite amount of affine
parameter the geodesic extends over infinite coordinate distances. In a more physical language one could 
reformulate the above statement by saying that it takes only a finite amount of affine parameter $\tau$ in 
order to reach the infinities of the incomplete space-time. As we will illustrate later in this 
chapter, the reason why the Randall-Sundrum II setup ceases to be geodesically complete is simply due to
a specific way of gluing two patches of $AdS_5$. 
One of the main motivations for this work was to provide an alternative to the Randall Sundrum II model that 
has the advantage of being geodesically complete while conserving the pleasant phenomenological features of 
the latter.
We divide the discussion of geodesics in two parts: in subsection \ref{RSgeod} we illustrate the effects of 
incomplete geodesics in the Randall-Sundrum II setup for timelike and lightlike geodesics. In the following
subsection \ref{ourgeod} we demonstrate why our setup is geodesically complete by looking at corresponding 
geodesics. 
Finally, we complement the discussions by illustrating the physics 
with the use of the Penrose-diagram of (the universal covering space-time of) $AdS_5$.
\subsection{Geodesics in the Randall-Sundrum II setup}
\label{RSgeod}

Our discussion of geodesics in this chapter is in no sense meant to be complete.
Without going into the details of the computations we merely intend to present the solutions of 
the geodesic equations in certain cases. For more general and more complete discussions of this issue we
refer to the literature, see e.g.~\cite{Muck:2000bb,Youm:2001qc} and references therein.
We first consider lightlike geodesics in the Randall-Sundrum II background metric given in appendix 
\ref{appParallel}, eq.~(\ref{RSmetric}). 
Let us suppose that a photon is emitted at the brane at $y=0$ in the positive $y$-direction at coordinate 
time $t=0$ then reflected at $y=y_1$ at the time $t=t_1$ to be observed by an 
observer on the brane at time $t=t_2$. In this situation $t$ corresponds to the proper time 
of an observer on the brane at rest. A simple calculation reveals
\begin{equation} \label{RSgeodint}
   t_2=2 t_1=\frac{2}{k} \left( e^{k y_1}-1\right).
\end{equation}
A brane bound observer will therefore note that it takes an infinite time for a photon to escape 
to $y=\infty$. However, parameterizing the same geodesic by an affine parameter $\tau$ reveals the
lightlike incompleteness of the Randall-Sundrum II space-time. Let the events of emission, refection (at
$y=y_1$) and arrival on the brane again be labeled by $\tau=0$, $\tau=\tau_1$ and $\tau=\tau_2$, 
respectively. Using the geodesic equation 
\begin{equation} \label{Geodesicequation}
  \frac{d^2 x^\mu}{d \tau^2}+\Gamma^\mu_{\nu \rho} \frac{d x^\nu}{d \tau}
  \frac{d x^\rho}{d\tau}=0 \, ,
\end{equation}
an easy computation shows:
\begin{equation} \label{RSgeodintau}
  \tau_2=2 \tau_1=\frac{2}{c k} \left(1- e^{-k y_1}\right).
\end{equation}
The constant $c$ is a remnant of the freedom in the choice of an affine parameter.\footnote{In general, two affine parameters $\lambda$ and $\tau$ are related by $\lambda=c \tau + d$, since 
this is the most general transformation leaving the geodesic equation (\ref{Geodesicequation}) invariant.
The choice of the origin of time to coincide with the emission of the photon only fixes $d$ but does not 
restrict $c$.}
From the last equation we see that now
\begin{equation} 
  \lim_{y_1 \to \infty} \tau_1 = \frac{1}{c k},
\end{equation}
meaning that in order to reach infinity ($y=\infty)$ in the extra dimension it takes only 
a finite amount $1/(c k)$ of affine parameter $\tau$, the expression of incompleteness 
of the Randall-Sundrum II space-time with respect to affinely parametrized lightlike geodesics.

In the case of timelike geodesics the inconsistency is even more striking. As shown in e.g. 
\cite{Gregory:2000rh}, a massive particle starting at the brane with vanishing initial velocity 
travels to $y=\infty$ in finite proper time given by $\tau_p=\pi/(2 k)$ while 
for a brane-bound observer this happens in an infinite coordinate time. 
To summarize, the Randall-Sundrum II brane-world model is geodesically incomplete both for null
and for timelike geodesics. As we will see in the next section, the geodesic incompleteness
is a direct consequence of the use of a particular coordinate system in $AdS_5$, the so-called 
Poincar\'e-coordinate system, which covers only a part of the full $AdS_5$ space-time.
We will also see that the problem of incomplete geodesics is absent in 
the setup we propose in this paper. 
  
\subsection{Geodesics in the background (\ref{metric}) and Penrose-diagram}
\label{ourgeod}

We would now like to answer similar questions to the ones considered in the previous section for the
background (\ref{metric}).
For example, we would like to know what time it takes for light to travel from the brane (at 
$\theta$=0) in the $\theta$-direction to a given point in the upper hemisphere with $\theta$ 
coordinate $\theta_1$, to be reflected and to return to the brane. As in the last section, 
$t$ and $\tau$ will denote the coordinate time (proper time of a stationary observer on the brane)
and the affine parameter used for parameterizing the geodesics, respectively. Again we choose 
$t=0$ ($\tau=0$) for the 
moment of emission, $t_1$ ($\tau_1$) for the reflection and $t_2$ ($\tau_2$) for the time where the photon 
returns to the brane. Due to the enormous hierarchy of distance scales in our model, one might wonder 
whether photons (or gravitons) are able to carry information from an arbitrary point on the brane to 
any other point on the brane connected to the first one by a null-geodesic in the extra dimension. 
For an observer on the brane such a possibility would be interpreted as 4-dimensional 
causality-violation. However, as we will see in the following, none of these possibilities exist in our model.
Omitting all details we find
\begin{equation} \label{Ourgeodint}
   t_2=2 t_1=4 R_U \coth\left(\frac{\omega\pi}{2}\right) 
   \arctan \left[\frac{\sinh( 
  \frac{\omega\theta_1}{2})}{\cosh\left[\frac{\omega}{2}\left(\pi-\theta_1\right)\right]} \right],
\end{equation}
so that an observer on the brane will see that the photon reaches the ``north pole'' $\theta_1=\pi/2$ at 
finite time
\begin{equation} 
  t_1=R_U \coth\left(\frac{\omega\pi}{2}\right) \arctan \left[ \sinh\left( \frac{\omega \pi}{2}\right)\right] 
  \approx R_U \frac{\pi}{2}\, .
\end{equation} 
Note that due to the warped geometry, it is $R_U$ entering the last relation and not $R$, so that 
even though the physical distance to the ``north pole'' is of the order of $R$ it takes a time of the 
order of $R_U$ for photons to reach it, excluding causality violation on the brane as 
discussed above. If the same geodesic is parametrized using an affine parameter we obtain 
\begin{equation} 
  \tau_2=2 \tau_1=\frac{2}{c}\frac{R}{\omega}\left[\tanh \left(\frac{\omega\pi}{2}\right)-
   \frac{\sinh \left[ \omega \left(\frac{\pi}{2}-\theta_1\right)\right]}{\cosh \frac{\omega\pi}{2}}\right]\, ,
\end{equation}
such that the amount of affine parameter needed to reach $\theta_1=\pi/2$ starting from the 
brane is:
\begin{equation} 
  \Delta \tau = \frac{1}{c}\frac{R}{\omega} \tanh \left(\frac{\omega\pi}{2}\right) \approx
  \frac{1}{c} \frac{R}{\omega} \, .
\end{equation}
Here again $c$ reflects the freedom in the choice of the affine parameter.

The results formally resemble those of the Randall-Sundrum II case. However, the important difference is that
in our case each geodesic can trivially be extended to arbitrary values of the affine parameter,
a simple consequence of the compactness of our space. Once the photon reaches the point $\theta=0$ it 
continues on its geodesic, approaching the brane, entering the southern ``hemisphere'', etc. 
It is clear that it needs an infinite amount of affine parameter in order to travel infinite coordinate 
distances. Therefore, the null geodesics in our setup which are the analogues of the incomplete 
null geodesics in the Randall-Sundrum II setup turn out to be perfectly complete due to the 
compactness of our space. The situation for timelike geodesics is fully analog to the case of the
null geodesics.

To end this section about the geometric properties of our model we would like to discuss the conformal 
structure of our space-time and point out differences to the Randall-Sundrum II setup. Let us review briefly 
the basic properties of $AdS_5$ space-time to the extend that we will need it in the following 
discussion.\footnote{We mainly follow \cite{Aharony:1999ti}.}
$AdS_5$ space-time can be thought of as the hyperboloid defined by 
\begin{equation} \label{hyperboloid}
  X_0^2+X_5^2-X_1^2-X_2^2-X_3^2-X_4^2=a^2
\end{equation}
embedded in a flat space with metric
\begin{equation} \label{AdSDefineMetric}
  ds^2=-dX_0^2-dX_5^2+dX_1^2+dX_2^2+dX_3^2+dX_4^2,
\end{equation}
$a$ being the so-called $AdS$-radius. The {\em global coordinates} of $AdS_5$ are defined by 
\begin{align}\label{GlobalCoordsDef}
  X_0&=a \cosh \chi \cos \tau,          &X_5=&a \cosh \chi \sin \tau, \nonumber \\
  X_1&=a \sinh \chi \cos \varphi_1,     &X_2=&a \sinh \chi \sin \varphi_1 \cos \varphi_2, \nonumber  \\
  X_3&=a \sinh \chi \sin \varphi_1 \sin \varphi_2 \cos \varphi_3,  
  &X_4=&a \sinh \chi \sin \varphi_1 \sin \varphi_2 \sin \varphi_3,
\end{align}
where the coordinates are confined by $0\leq \chi$, $-\pi \leq \tau \leq \pi$, $0\leq\varphi_1\leq \pi$,
$0\leq\varphi_2\leq \pi$, $0\leq\varphi_3\leq 2\pi$ and $\tau=-\pi$ is identified with $\tau=\pi$. 
These coordinates cover the full hyperboloid exactly once. Allowing $\tau$ to take values on the real line
without the above identification of points gives the universal covering space $CAdS_5$ of $AdS_5$.\footnote{Whenever we used the word $AdS_5$ so far in this paper we actually meant $CAdS_5$. 
For reasons of clarity, however, we will brake with this common practice 
in the rest of this section.}
In these coordinates, the line element (\ref{AdSDefineMetric}) can be written:\footnote{\label{changecoords}It is obvious that the metric (\ref{metric}) for $0\leq \theta \leq \pi/2$ 
($-\pi/2 \leq \theta \leq 0$) reduces to the above line-element (\ref{AdSGlobalCoords}) under the 
following coordinate transformation: $\chi=\omega (\frac{\pi}{2}\mp \theta)$, 
$\tau= t \omega/\left[R \cosh \left(\frac{\omega\pi}{2}\right)\right]$, together with $a=R/\omega$.}
\begin{equation} \label{AdSGlobalCoords}
  ds^2=a^2\left(-\cosh^2\chi \; d\tau^2+d\chi^2+\sinh^2\chi \; d\Omega_3^2\right) \, .
\end{equation}

Another coordinate system can be defined by 
\begin{align}\label{PoincareCoordsDef}
  X_0&= \frac{1}{2 u} \left[1+u^2\left(a^2+\vec{x}^2-\bar{t}^{\,2}\right)\right] ,
 &X_5=&a \, u \, \bar{t} , \nonumber \\
  X^i&=a \, u \, x^i, \; i=1,2,3\; ,  
 &X^4=& \frac{1}{2 u} \left[1-u^2\left(a^2-\vec{x}^2+\bar{t}^{\, 2}\right)\right],
\end{align}
with $u>0$, $\bar{t} \in (-\infty, \infty)$ and $x^i \in (-\infty, \infty)$.
In these {\em Poincar\'e coordinates} the line element (\ref{AdSDefineMetric}) takes the form
\begin{equation} \label{AdsPoincareCoords}
  ds^2= a^2 \left[ \frac{du^2}{u^2} + u^2 \left(-d\bar{t}^{\, 2}+d\vec{x}^2\right)\right].
\end{equation}
In contrast to the global coordinates, the Poincar\'e coordinates do not cover the whole of 
the $AdS_5$ and $CAdS_5$ space-times \cite{Aharony:1999ti}. From (\ref{AdsPoincareCoords}), after changing coordinates 
according to $dy=-a \, du/u$ and rescaling $t$ and $x^i$ by the $AdS$-radius $a$ we recover the original 
Randall-Sundrum II coordinate system given in (\ref{RSmetric}). 
The restrictions on $y$ in the Randall-Sundrum II setup further 
limit the range covered by their coordinate system to the $0<u \leq 1$ domain of the Poincar\'e coordinates.

Coming back to the global coordinates, we introduce $\rho$ by
\begin{equation} 
  \tan \rho = \sinh \chi \;\; \mbox{with} \;\; 0\leq \rho < \frac{\pi}{2},
\end{equation}
so that (\ref{AdSGlobalCoords}) becomes 
\begin{equation} \label{PenroseCoords}
  ds^2=\frac{a^2}{\cos^2 \rho} \left(-d\tau^2+d\rho^2+\sin^2\rho \; d\Omega_3^2 \right).
\end{equation}

\begin{figure}[htbp]
\begin{center}
\input{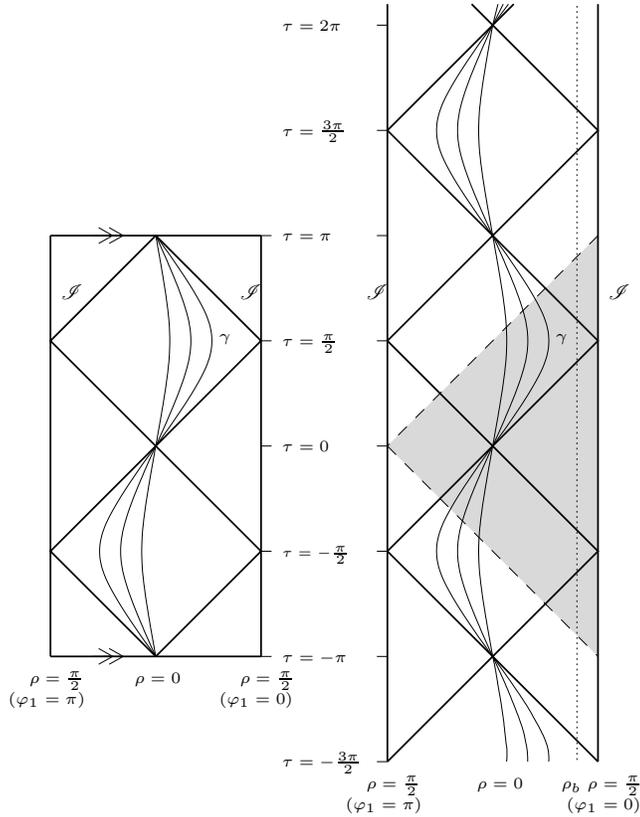}
\end{center}
\caption{Penrose diagram of $AdS_5$ space and its universal covering space $CAdS_5$}
\label{penrose}
\end{figure}

The Penrose-diagrams of $AdS_5$ space-time and its universal covering space-time $CAdS_5$ are shown in 
Fig.~\ref{penrose}, see \cite{HawkingEllis,Avis:1977yn}.
While the $AdS_5$ space-time contains closed timelike curves (denoted by $\gamma$ in the figure), 
its universal covering space-time $CAdS_5$ does not.
The arrows in the left diagram indicate that the lines $\tau=-\pi$ and $\tau=\pi$ should be identified.
The symbol $\mathscr{I}$ stands for the timelike surface $\rho=\pi/2$ (spatial infinity). It is this 
surface which is responsible for the absence of a Cauchy-surface in $AdS$-space. 
We will concentrate in the following 
on the $CAdS_5$ diagram. First we note that each point in the diagram corresponds to a $3$-sphere.
The shaded region indicates the patch covered by the Poincar\'e (and Randall-Sundrum) coordinates.
Note that the position of the Randall-Sundrum brane cannot be represented in a simple way in the
Penrose diagram of $CAdS_5$. The reason is that the $u=\mbox{const.}$ hypersurfaces of 
the Poincar\'e coordinates 
generate a slicing of flat $4$-dimensional Minkowski space-times while the points in the diagram represent
(curved) $3$-spheres.
From the Penrose diagram it is immediately clear that the Randall-Sundrum-II space-time is geodesically 
incomplete. The timelike curves denoted by $\gamma$ emanating from the origin ($\rho=0,\tau=0$) 
will all eventually exit the shaded region after a finite coordinate time $\tau$. These geodesics appear
inextendible from the point of view of the Randall-Sundrum space-time. The problems arise due to the 
arbitrary cutting of a space-time along the borders of a given coordinate patch which covers only a part of 
the initial space-time.
Similar conclusions can be drawn for null geodesics, represented by lines making angles of $45$ degrees 
with the vertical lines in Fig.~\ref{penrose}.
The vertical dotted line with topology $\mathbb{R} \times S^3$ at coordinate 
$\rho_b=\arctan \left[ \sinh \left(\frac{\omega\pi}{2} \right)\right]$ close to $\rho=\pi/2$ in the right diagram 
corresponds to the location of our curved three brane, $\tau$ being 
proportional to our time coordinate $t$ (see footnote \ref{changecoords}). We note that since 
$\omega > 50$, this line is drawn by far
too distant from $\rho=\pi/2$ as a simple expansion shows:
\begin{equation} 
  \rho_b=\arctan \left[ \sinh \left(\frac{\omega \pi}{2}\right)\right] \sim \frac{\pi}{2} -2 e^{-\frac{\omega\pi}{2}}+ 
  \mathcal{O}(e^{-\frac{3 \omega\pi}{2}}) \, .
\end{equation}
The slice of $CAdS_5$ to the right of $\rho_b$ 
($\rho_b \leq \rho \leq \pi/2$) is discarded and replaced by another copy of the slice to the left 
of $\rho_b$ ($0\leq\rho\leq \rho_b$).

From the Penrose diagram we can also deduce that our space-time is geodesically complete. Timelike 
geodesics (emanating from ($\rho=0,\tau=0$)) cross the brane $\rho_b$ at some point or return to the origin
depending on the initial velocities of the particles that define them. In both cases (as in the case of 
null geodesics) there is no obstacle to extend the affine parameter to larger and larger values. 

As we already mentioned, the absence of a Cauchy surface in $CAdS$-space-time is due to the existence 
of the timelike surface $\mathscr{I}$ at spatial infinity. By construction, our space-time excludes 
$\mathscr{I}$ so that the natural question arises whether it is legitimate to revert the above argument 
and conclude the existence of a Cauchy surface in our model. Though interesting, we do not intend to 
elaborate this question any further in this paper.

\section{Gravity localization on the brane}
\setcounter{equation}{0}
\label{three}

There are at least two equivalent ways of addressing the problem of the localization of gravity
in brane-world scenarios. Both ways have their advantages and disadvantages. The first way is
based on a detailed study of the Kaluza-Klein excitations of the graviton. After integrating out the 
extra dimension(s) one obtains an effective 4 dimensional Lagrangian involving the full tower of 
Kaluza-Klein gravitons. Considering the low energy scattering process of two test particles on the brane via 
exchange of Kaluza-Klein excitations allows to relate the non-relativistic scattering amplitude to the
static potential felt by the two test particles. Since the coupling of each individual Kaluza-Klein 
particle to matter on the brane is proportional to the value of its transverse wave-function at the position
of the brane, this approach necessitates a proper normalization of all Kaluza-Klein modes. An advantage of
this approach is the possibility of distinguishing between the contributions to the potential coming from 
the zero mode (Newton's law) and from the higher modes (corrections thereof).

The second approach is based on a direct calculation of the graviton two-point function in the space-time
under consideration, having the obvious advantage of bypassing all technicalities related to the Kaluza-Klein
spectrum and the wave-function normalization. However, a physical interpretation of the effects of individual 
Kaluza-Klein modes from the point of view of a 4-dimensional observer is, to say the least, not 
straightforward. Finally, due to the equivalence of the two approaches it is clearly possible to fill this 
gap by reading of the Kaluza-Klein spectrum and the (normalized) wave-functions from the two-point function 
by locating its poles and determining corresponding residues.   

In the setup considered in this paper we choose to work in the second approach, the direct evaluation of 
the Green's functions, due to additional difficulties arising from the non-Minkowskian nature of the 
induced metric on the brane. The lack of Poincar\'e invariance on the brane clearly implies the absence of 
this symmetry in the effective 4-dimensional Lagrangian as well as non-Minkowskian dispersion relations for 
the Kaluza-Klein modes. 

\subsection{Perturbation equations and junction conditions}

We would like to study the fluctuations $H_{ij}$ of the metric in the background (\ref{metric}) 
defined by:
\begin{equation} 
  ds^2=-\sigma^2 dt^2+R_U^2 \gamma^2 \left[\eta_{ij}+2 H_{ij}\right] d\varphi^i d\varphi^j +R^2 d\theta^2.
\end{equation}
$H_{ij}$ transform as a transverse, traceless second-rank tensor with respect to coordinate transformations
on the maximally symmetric space with metric $\eta_{ij}$, a $3$-sphere in our case:
\begin{equation} 
  \eta^{ij} H_{ij}=0, \quad \eta^{i j} \tilde{\nabla}_i H_{j k}=0,
\end{equation}
where $\tilde{\nabla}$ denotes the covariant derivative associated with the metric $\eta_{ij}$ on $S^3$.
The gauge invariant symmetric tensor $H_{ij}$ does not couple to the vector and scalar perturbations. 
Its perturbation
equation in the bulk is obtained from the transverse traceless 
component of the 
perturbed Einstein equations in $5$ dimensions:
\begin{equation} 
  \delta R_{M N}-\frac{2}{3} \Lambda \delta g_{M N}=0 \, .
\end{equation}
Specifying this equation to the case of interest of a static background we  find:
\begin{equation} \label{Hfluct}
  \frac{1}{\sigma^2} \ddot{H}-\frac{1}{R_U^2\gamma^2}\tilde{\Delta} H_{ij} -
  \frac{1}{R^2}H_{ij}^{\prime \prime}-
  \frac{1}{R^2} \left( \frac{\sigma^\prime}{\sigma}  
  +3\frac{\gamma^\prime}{\gamma}\right)H_{ij}^{\prime}
  +\frac{2}{R_U^2\gamma^2}H_{ij} = 0,
\end{equation}
where $\tilde{\Delta}$ denotes the Laplacian on $S^3$. The last equation is valid in the bulk in the 
absence of sources. We can now conveniently expand $H_{ij}$ in the basis of the symmetric transverse 
traceless tensor harmonics $\hat{T}^{(l \lambda)}_{ij}$ on $S^3$ 
\cite{Rubin:tc,Allen:1986tt,Mukohyama:2000ui}:
\begin{equation} \label{Hexpand}
  H_{ij}=\sum_{l=2}^\infty \sum_\lambda \Phi^{(l \lambda)}(t,\theta) \hat{T}^{(l \lambda)}_{ij},
\end{equation}
where the $\hat{T}^{(l \lambda)}_{ij}$ satisfy
\begin{align} \label{TTHarmonics}
  \tilde{\Delta} \hat{T}^{(l \lambda)}_{ij} + k_l^2 \hat{T}^{(l \lambda)}_{ij} =0, 
  \qquad k_l^2= l(l+2)-2, \qquad l=2,3,... \nonumber  \\
  \eta^{ij}\tilde{\nabla}_i \hat{T}^{(l \lambda)}_{jk}=0, \qquad 
  \eta^{ij} \hat{T}^{(l \lambda)}_{ij} =0, \qquad 
  \hat{T}^{(l \lambda)}_{[ij]}=0.
\end{align}
The sum over $\lambda$ is symbolic and replaces all eigenvalues needed to describe the 
full degeneracy of the subspace of solutions for a given value of $l$.
Introducing the expansion (\ref{Hexpand}) into (\ref{Hfluct}) and using the orthogonality 
relation of the tensor harmonics \cite{Allen:1986tt} 
\begin{equation} \label{TTTOrthogonality}
  \int \sqrt{\eta } \, \eta^{ik} \, \eta^{jl} \, \hat{T}^{(l \lambda)}_{ij} 
   \, \hat{T}^{(l' \lambda')}_{kl} d^3\varphi = 
   \delta^{l l'} \, \delta^{\lambda \lambda'} \, ,
\end{equation}
we obtain:
\begin{equation} \label{Hfluctexpansion}
  \frac{1}{\sigma^2} \ddot{\Phi}^{(l \lambda)}- \frac{1}{R^2}\Phi^{(l \lambda) \prime \prime}-
  \frac{1}{R^2} \left( \frac{\sigma^\prime}{\sigma} + 3 \frac{\gamma^\prime}{\gamma}\right)
   \Phi^{(l \lambda) \prime}+\frac{l(l+2)}{R_U^2\gamma^2} \Phi^{(l \lambda)}=0.
\end{equation}
This equation has to be compared to the equation of a massless scalar field in the background 
(\ref{metric}). After expanding the massless scalar in the corresponding scalar harmonics on
$S^3$ we recover (\ref{Hfluctexpansion}) with the only difference in the eigenvalue parameters
due to different spectra of the Laplacian $\tilde{\Delta}$ for scalars and for tensors.
Motivated by this last observation and in order to avoid the technicalities related to the tensorial nature
of the graviton $H_{ij}$ we confine ourselves in this paper to the study of the Green's functions 
of a massless scalar field in the background (\ref{metric}).

The differential equation (\ref{Hfluctexpansion}) alone does not determine $H_{ij}$
uniquely. We have to impose proper boundary conditions for $H_{ij}$. While imposing square 
integrability will constitute one boundary condition at $\theta=\pm \pi/2$, 
the behavior of $H_{ij}$ on the brane will be dictated by the Israel junction condition \cite{Israel:rt}:
\begin{equation} \label{israelcondition}
  : K_{\mu \nu} : \;\;\; =-\frac{8 \pi}{M^3}\left(T_{\mu\nu}-\frac{1}{3} T_\kappa^{\;\kappa} \bar{g}_{\mu \nu}
  \right).
\end{equation}
Here $K_{\mu\nu}$ and $\bar{g}_{\mu\nu}$ denote the extrinsic curvature and the induced metric on the brane, 
respectively. $T_{\mu\nu}$ is the $4$-dimensional stress-energy tensor on the 
brane\footnote{\; $T^0_{\;\;0}=c_0$, and $T^i_{\;\;j}=\delta^i_{\;\;j} c$ in the notation of section 
(\ref{SingularBrane}).} and the symbol $: \ldots :$ is defined in (\ref{::def}).  
In our coordinate system the extrinsic curvature is given by:
\begin{equation} \label{extrCurvature}
  K_{\mu\nu}=\frac{1}{2 R} \frac{\partial{\bar{g}_{\mu \nu}}}{\partial \theta} \, .
\end{equation}
Hence, the non-trivial components of the Israel condition become:
\begin{align} 
  : \frac{\sigma^\prime}{\sigma} : &= -\frac{8 \pi R}{M^3} \left(\frac{2}{3} c_0-c 
     \right) \, , \label{Israel00BG} \\
  : \frac{\gamma^\prime}{\gamma} \left( \eta_{ij}+2 H_{ij} \right) + H^\prime_{ij} : 
   &= \frac{8 \pi R}{3 M^3} c_0 \left(\eta_{ij}+2 H_{ij} \right). \label{Israelij}
\end{align}
Separating the background from the fluctuation in (\ref{Israelij}) we find:
\begin{align} 
  : \frac{\gamma^\prime}{\gamma} : &= \frac{8 \pi R}{3 M^3} c_0 \, , \label{IsraelijBG}\\
  : H^\prime_{ij} : &= 0 \label{IsraelijFluct} \, ,
\end{align}
where we only used the continuity of $H_{ij}$ on the brane. While eq.~(\ref{IsraelijBG}) directly coincides 
with eq.~(\ref{FineTuneI}) of section (\ref{SingularBrane}), eq.~(\ref{Israel00BG}) turns out 
to be a linear combination of (\ref{FineTuneI}) and (\ref{FineTuneII}).

Eq.~(\ref{IsraelijFluct}) taken alone implies the continuity of $H^\prime_{ij}$ on the brane. If 
in addition the fluctuations are supposed to satisfy the $\theta \to-\theta$ symmetry, this condition
reduces to a Neumann condition, as for example in our treatment of the scalar two-point function 
in the Randall-Sundrum II case (see appendix \ref{appParallel}).

In the next section we are going to find the static Green's functions of a massless scalar field in the 
Einstein static universe background. This will serve as a preparatory step for how to handle the more 
complicated case of a non-invertible Laplacian. More importantly, it will provide the necessary reference
needed for the interpretation of the effect of the extra dimension on the potential between two test masses 
on the brane.

\subsection{Gravity in the Einstein static universe}
As announced in the previous section, we will concentrate on the massless scalar field. 
Our aim is to solve the analog of Poisson equation in the Einstein static universe background.
Due to its topology $\mathbb{R} \times S^3$ we will encounter a difficulty related to the existence 
of a zero eigenvalue of the scalar Laplacian on $S^3$ necessitating the introduction of a modified 
Green's function.\footnote{For an elementary introduction to the concept of modified Green's functions see e.g
\cite{Stakgold}.} Note that in the more physical case where the full tensor structure of the 
graviton is maintained no such step is necessary since all eigenvalues of the tensorial Laplacian
are strictly negative on $S^3$, see (\ref{TTHarmonics}).

\subsubsection{Definition of a modified Green's function}
We take the metric of Einstein's static universe in the form:
\begin{equation} \label{esu_metric}
  ds^2=g_{\mu\nu}dx^\mu dx^\nu = - dt^2+A^2 d \Omega_3^2
\end{equation}
with $d\Omega_3^2$ being the line element of a $3$-sphere given in (\ref{S3metric}) and $A$ being its 
constant radius. 
The equation of a massless scalar field then becomes:
\begin{equation} \label{ScaFieldEqESU}
  \frac{1}{\sqrt{-g}}\partial_\mu \left[\sqrt{-g} \, g^{\mu\nu}\partial_\nu u(t,\vec{x}) \right]= j(t,\vec x)
\end{equation}
which in the static case reduces to
\begin{equation} \label{StaticEqESU}
  \mathcal{D} u(\vec x) \equiv \frac{1}{A^2} \tilde{\Delta} u(\vec x) = j(\vec x) \, ,
\end{equation}
where $\tilde{\Delta}$ is the scalar Laplacian on $S^3$ and we choose to write $\vec x$ for the collection
of the three angles on $S^3$.
All functions involved are supposed to obey periodic boundary condition on $S^3$ so that in
Green's identity all boundary terms vanish:
\begin{equation} \label{GreensIdentityESU}
  \int \sqrt{-g} \, \left(\mathcal{D} \, v(\vec x)\right) \, \overline{u}(\vec x) \, d^3 \vec{x} = 
  \int \sqrt{-g} \, v(\vec x) \, \overline{\mathcal{D} \, u(\vec x)} \, d^3 \vec{x}=
  \int \sqrt{-g} \, v(\vec x) \, \overline{j}(\vec x) \, d^3\vec{x} \, .
\end{equation}
Here and in the following bars denote complex conjugate quantities.
The operator $\mathcal{D}$ trivially allows for an eigenfunction with zero eigenvalue (the constant
function on $S^3$):
\begin{equation}
  \mathcal{D} u_0(\vec x) = 0 \, ,\quad \int \sqrt{-g} \, u_0(\vec x) \, \overline{u}_0(\vec x) \, 
  d^3 \vec x=1 \, .
\end{equation}  
Then, since the homogeneous equation has not only the trivial solution ($u=0$) which satisfies 
the periodic boundary conditions in the angular variables, the operator $\mathcal{D}$ cannot be 
invertible. Therefore, there
is no solution to the equation (\ref{StaticEqESU}) for an arbitrary source.  
In order to still define a ``modified'' Green's function we have to restrict the possible 
sources to sources that satisfy the following solvability condition:
\begin{equation} \label{solvabilityESU}
  \int \sqrt{-g} \, j(\vec x) \, \bar{u}_0(\vec x) \, d^3\vec x=0.
\end{equation}
Formally this condition can be obtained by replacing $v(\vec x)$ by the non-trivial solution 
of the homogeneous equation $u_0(\vec x)$  in (\ref{GreensIdentityESU}).
We now define the modified Green's function as a solution of
\begin{equation} \label{ModGreenFunctionESU}
  \mathcal{D}_x \mathcal{G}(\vec x,\vec x')=\frac{\delta^3(\vec x-\vec x')}{\sqrt{-g}}-u_0(\vec x)
  \, \bar{u}_0(\vec x')  \, .
\end{equation}
Replacing now $v(\vec x)$ by $\mathcal{G}(\vec x,\vec x')$ in Green's identity (\ref{GreensIdentityESU}) we obtain (after complex conjugation) the desired integral representation for the solution of (\ref{StaticEqESU}):
\begin{eqnarray} \label{esuIntReprESU}
  u(\vec x) = C u_0(\vec x)+\int \sqrt{-g} \, j(\vec x') \, \overline{\mathcal{G}}(\vec x',\vec x) \, 
  d^3 \vec x' \, ,
\end{eqnarray}
where $C$ is given by
\begin{equation} 
  C=\int \sqrt{-g} \, u(\vec x') \, \overline{u}_0(\vec x') \, d^3\vec x' \, .
\end{equation}
Note that the source 
\begin{equation}
  j_0(\vec x)=\frac{\delta^3(\vec x-\vec x')}{\sqrt{-g}}-u_0(\vec x)\bar{u}_0(\vec x')
\end{equation}
trivially satisfies the solvability condition (\ref{solvabilityESU}) and represents a point source at 
the location $\vec x=\vec x'$ compensated by a uniform negative mass density. To fix the 
normalization  of $u_0(\vec x)$ we write:
\begin{equation}
  u_0(\vec x)=N_0 \Phi_{100}(\vec x) \,\,\, \mbox{with} \,\,\, \int \sqrt{\eta} \, \Phi_{100}(\vec x) 
  \overline{\Phi}_{100}(\vec x) d^3 \vec x=1,
\end{equation}
where $\Phi_{100}=\frac{1}{\pi \sqrt{2}}$ and so
\begin{equation}
  \int \sqrt{-g} \, u_0(\vec x) \, \overline{u}_0(\vec x) \, d^3\vec x = \vert N_0 \vert^2 A^3 
  \underbrace{\int \sqrt{\eta} \, \Phi_{100}(\vec x) \, \overline{\Phi}_{100}(\vec x) \, d^3 \vec x}_{=1} = 
 \vert N_0 \vert^2 A^3=1 \, ,
\end{equation}
so that
\begin{equation}
  u_0(\vec x)=\frac{1}{A^{3/2}}\Phi_{100}(\vec x).
\end{equation}
In order to solve the differential equation (\ref{ModGreenFunctionESU}) defining the modified Green's 
function $\mathcal{G}(\vec x, \vec x')$, we expand in eigenfunctions of the Laplace operator on $S^3$, 
the so-called scalar harmonics $\Phi_{\lambda l m}$ with properties 
(see e.g.~\cite{Kodama:2000fa,GribMamaMost}):
\begin{equation} \label{ScalarHarmonicsESU}
  \tilde{\Delta} \Phi_{\lambda l m} =(1-\lambda^2 )\Phi_{\lambda l m} 
  \qquad \lambda=1,2,\ldots; l=0,\ldots,\lambda-1;m=-l,\ldots,l \, .
\end{equation} 
Our ansatz therefore reads:
\begin{equation}
  \mathcal{G}(\vec x, \vec x')=\sum_{\lambda=1}^\infty \sum_{l=0}^{\lambda-1}\sum_{m=-l}^l 
  \Phi_{\lambda l m}(\vec x) \, c_{\lambda l m}(\vec x') \, .
\end{equation}
Introducing this in (\ref{ModGreenFunctionESU}) we obtain 
\begin{equation} 
  \sum_{\lambda=1}^{\infty}\sum_{l=0}^{\lambda-1}\sum_{m=-l}^l \left[-\frac{\lambda^2-1}{A^2}
  c_{\lambda l m}(\vec x')\right] \Phi_{\lambda l m}(\vec x)= 
   \frac{\delta^3(\vec x-\vec x')}{\sqrt{-g}}-u_0(\vec x)\, \overline{u}_0(\vec x')\, .
\end{equation} 
If we now multiply by $\Phi_{\lambda' l' m'}(\vec x) \, \sqrt{\eta}$ and integrate over $S^3$ we find
\begin{equation}
 -\frac{\lambda'^2-1}{A^2}c_{\lambda' l' m'}(\vec x') =\frac{1}{A^3} 
  \overline{\Phi}_{\lambda' l' m'}(\vec x')-\bar{u}_0(\vec x') \int\limits_{S^3} \sqrt{\eta} \,  
  u_0(\vec x) \, \overline{\Phi}_{\lambda' l' m'}(\vec x) \, d^3 \vec x,
\end{equation}
where we made use of the orthogonality relation of the scalar harmonics
\begin{equation} \label{Harmonicsorthogonality}
 \int \sqrt{\eta} \, \overline{\Phi}_{\lambda l m}(\vec x) \, \Phi_{\lambda' l' m'}(\vec x) \, d^3 \vec x=
 \delta_{\lambda \lambda'}   \delta_{l l'} \delta_{m m'}.
\end{equation}
In the case $\lambda'=1$ (and vanishing $l'$ and $m'$) the above equation is identically
satisfied for all values of $c_{1 0 0}(\vec x')$.\footnote{This means that we have the freedom to choose 
$c_{1 0 0}(\vec x')$ freely. Our choice is $c_{1 0 0}(\vec x')=0$ without restricting generality 
since from eq. (\ref{esuIntReprESU}) we immediately conclude that for sources satisfying 
(\ref{solvabilityESU}) there will never be any contribution to $u(\vec x)$ coming from $c_{1 0 0}(\vec x')$.}

In the case $\lambda'\neq 1$ the coefficient $c_{\lambda' l' m'}(\vec x')$ follows to be
\begin{equation}
 c_{\lambda' l' m'}(\vec x')=\frac{\overline{\Phi}_{\lambda' l' m'}(\vec x')}{A(1-\lambda'^2)}
\end{equation}
so that the formal solution for the modified Green's function can be written as
\begin{equation} 
  \mathcal{G}(\vec x,\vec x') =
  \sum_{\lambda=2}^{\infty} \sum_{l=0}^{\lambda-1} \sum_{m=-l}^{l} 
   \frac{\Phi_{\lambda l m}(\vec x) \bar{\Phi}_{\lambda l m}(\vec x')}{A(1-\lambda^2)}\, .
\end{equation}
Due to the maximal symmetry of the $3$-sphere, the Green's function 
$\mathcal{G}(\vec x,\vec x')$ can only depend on 
the geodesic distance $s(\vec x, \vec x') \in [0,\pi]$ between the two points 
$\vec x$ and $\vec x'$:
\begin{eqnarray} \label{DistanceOnS3}
  \cos s &=& \cos \varphi_1 \cos \varphi'_1+\sin\varphi_1\sin\varphi'_1 \cos\beta \, ,\nonumber\\
  \cos\beta&=& \cos\varphi_2 \cos \varphi'_2+ \sin\varphi_2\sin\varphi'_2 \cos(\varphi_3-\varphi'_3) \, ,
\end{eqnarray}
which in the case $\varphi_2=\varphi'_2$ and $\varphi_3=\varphi'_3$ clearly reduces to 
$s=\varphi_1-\varphi'_1$. Indeed, the sum over $l$ and $m$ can be performed 
using\footnote{See e.g.~\cite{GribMamaMost}.}
\begin{equation} \label{SumOverS3}
  \sum_{l=0}^{\lambda-1} \sum_{m=-l}^{l} \bar{\Phi}_{\lambda l m}(\vec x) 
   \Phi_{\lambda l m}(\vec x')=\frac{\lambda}{2 \pi^2} \frac{\sin \left(\lambda s\right)}{\sin s}
\end{equation}
such that even the remaining sum over $\lambda$ can be done analytically:
\begin{equation} \label{ESUresult}
  \tilde{\mathcal{G}}(s) \equiv \mathcal{G}(\vec x,\vec x')=
   \frac{1}{8 \pi^2 A}-\frac{1}{4\pi A}\left[(1-\frac{s}{\pi})\cot s \right]\, , \qquad s \in [0,\pi]\, .
\end{equation}
To interpret this result, we develop $\tilde{\mathcal{G}}(s)$ around $s=0$ 
obtaining\footnote{Note that the constant term in the expansion (\ref{expansioninsESU}) is specific 
to our choice of $c_{1 0 0}(\vec x')$.}
\begin{equation} \label{expansioninsESU}
  \tilde{\mathcal{G}}(s)=\frac{1}{A}\left[-\frac{1}{4\pi s} +\frac{3}{8\pi^2}+ \frac{s}{12 \pi}
  +\mathcal{O}(s^2) \right] \, .
\end{equation}
By introducing the variable $r=s A$ and by treating $\tilde{\mathcal{G}}(s)$ as a
gravitational potential we find 
\begin{equation}
  \frac{1}{A} \frac{d\tilde{\mathcal{G}}(s)}{ds}=\frac{1}{4\pi r^2}+\frac{1}{12\pi A^2}+\mathcal{O}(r/A^3).
\end{equation}
We notice that for short distances, $r \ll A$, we find the expected flat result whereas the corrections to 
Newton's law become important at distances $r$ of the order of $A$ in the form of a constant attracting
force.\footnote{There exists numerous articles treating the gravitational potential of a point source 
in Einstein's static universe (see \cite{Astefanesei:2001cx,Nolan:1999wf} and references therein). 
We only would like to point out here similarities between our  
result (\ref{ESUresult}) and the line element of a Schwarzschild metric in an Einstein static universe
background given in \cite{Astefanesei:2001cx}.}

\subsection{Modified Green's function of a massless scalar field in the background space-time (\ref{metric})}
We are now prepared to address the main problem of this work namely the computation
of the modified Green's function of a massless scalar field in the background space-time
(\ref{metric}). Since our main interest focuses again on the low energy properties of the two-point
function, we will limit ourselves to the static case. The main 
line of reasoning is the same as in the previous section. Due to the fact that our space has the
global topology of a 4-sphere $S^4$, we again are confronted with a non-invertible differential operator.
We would like to solve 
\begin{equation} \label{ScalarEq}
  \mathcal{D} u(\varphi_i,\theta) 
  \equiv \frac{\tilde{\Delta} u(\varphi_i,\theta)}{R_U^2 \gamma(\theta)^2}+
  \frac{1}{R^2} \frac{1}{\sigma(\theta)\gamma(\theta)^3} \frac{\partial}{\partial\theta}
  \left[ \sigma(\theta)\gamma(\theta)^3 \frac{\partial u(\varphi_i,\theta)}{\partial \theta}\right]=
  j(\varphi_i,\theta)  \, ,
\end{equation}
where the independent angular variable ranges are  
$0\leq\varphi_1\leq\pi;\,0\leq\varphi_2\leq\pi;\, 0\leq\varphi_3\leq 2 \pi;
 \,-\pi/2\leq\theta\leq\pi/2$. 
Every discussion of Green's functions is based on Green's identity relating the differential 
operator under consideration to its adjoint operator. Since $\mathcal{D}$ is formally self-adjoint we have
\begin{eqnarray} \label{GreensIdentity}
  \int\limits_0^\pi d\varphi_1 \int\limits_0^\pi d\varphi_2 \int\limits_0^{2\pi} d\varphi_3 
  \int\limits_{-\frac{\pi}{2}}^{\frac{\pi}{2}} d\theta \sqrt{-g}
  \left[(\mathcal{D} v) \overline{u} -v \overline{(\mathcal{D} u)}\right]=\nonumber \\
  \int\limits_0^\pi d\varphi_1 \int\limits_0^\pi d\varphi_2 \int\limits_0^{2\pi} d\varphi_3 \frac{R_U^3}{R} 
  \sqrt{\eta}\left[ \sigma(\theta)\gamma(\theta)^3 
  \left( \bar{u} \frac{\partial v}{\partial\theta}-
     \frac{\partial \bar{u}}{\partial\theta} v \right) 
   \right]_{-\frac{\pi}{2}}^{\phantom{-}\frac{\pi}{2}}+\ldots \, ,
\end{eqnarray}
where we dropped the arguments of $u$ and $v$ for simplicity.
The dots in (\ref{GreensIdentity}) refer to boundary terms in the variables $\varphi_i$ and since 
we again employ an eigenfunction expansion in scalar harmonics on $S^3$, these 
boundary terms will
vanish. In order to find an integral representation of the solution $u(\varphi_i,\theta)$ of 
(\ref{ScalarEq}) we have to impose appropriate boundary conditions on $u$ and $v$ at 
$\theta=\pm \pi/2$. For the time being we assume this to be the case such that all boundary
terms in (\ref{GreensIdentity}) vanish and proceed with the formal solution of (\ref{ScalarEq}).
We will address the issue of boundary conditions in $\theta$ in detail in appendices \ref{appParallel} and
\ref{appDEQ}.

In the following we will collectively use $x$ instead of $(\varphi_i, \theta)$.
The homogeneous equation $\mathcal{D} u(x)=~0$ does not have a unique solution under the 
assumption of periodic boundary conditions. In addition to the trivial solution $(u=0)$ we also find 
\begin{equation} \label{ZeroMode}
  \mathcal{D} \, u_0(x) = 0\, , \qquad \int \sqrt{-g} \, u_0(x) \, 
   \overline{u}_0(x) \, d^4x=1 \, .
\end{equation} 
Therefore, the corresponding inhomogeneous equation (\ref{ScalarEq}) does not have a solution unless
we again restrict the space of allowed sources: 
\begin{equation} \label{solvability}
  \int \sqrt{-g} \, j(x) \, \overline{u}_0(x) \, d^4x=0.
\end{equation}
As in the last section, this condition can be obtained by replacing $v$ by the non-trivial solution 
of the homogeneous equation $u_0$  in (\ref{GreensIdentity}). 
We now define the modified Green's function by
\begin{equation} \label{ModGreenFunction}
  \mathcal{D}_x \mathcal{G}(x,x')=\frac{\delta^4(x-x')}{\sqrt{-g}}-u_0(x)\overline{u}_0(x') \, .
\end{equation}
From Green's identity (\ref{GreensIdentity}), with $v(x)$ given by $\mathcal{G}(x,x')$, 
we again obtain after complex conjugation the desired integral representation:
\begin{eqnarray} \label{esuIntRepr}
  u(x) = C u_0(x)+\int \sqrt{-g} \, j(x') \overline{\mathcal{G}(x',x)} \, d^4x' \, ,
\end{eqnarray}
with 
\begin{equation} 
  C=\int \sqrt{-g} \, u(x') \, \overline{u}_0(x') \, d^4x' \, .
\end{equation}
As before, the source
\begin{equation}
  j_0(x)=\frac{\delta^4(x-x')}{\sqrt{-g}}-u_0(x)\, \overline{u}_0(x')
\end{equation}
satisfies the solvability condition (\ref{solvability}) by construction.
The normalization of the constant mode $u_0(x)$ is slightly more involved than 
before due to the nontrivial measure $\sigma(\theta)\gamma(\theta)^3$ in the $\theta$ 
integration. By inserting  
\begin{equation} 
  u_0(x)=N_0 \Phi_{1 0 0}(\varphi_i) \chi_1(\theta) \, \, \mbox{with}\,\, \chi_1(\theta)=1
\end{equation}
in the integral in (\ref{ZeroMode}) we obtain 
\begin{equation}
  N_0 =\left[\frac{2\omega}{R R_U^3 \tanh \left(\frac{\omega \pi}{2}\right)}\right]^{\frac{1}{2}} \,.
\end{equation}
For the solution of eq. (\ref{ModGreenFunction}) we use the ansatz
\begin{equation}
  \mathcal{G}(\varphi_i,\varphi'_i,\theta,\theta')=\sum_{\lambda=1}^{\infty}\sum_{l=0}^{\lambda-1}
  \sum_{m=-l}^l \Phi_{\lambda l m}(\varphi_i) \, c_{\lambda l m}(\varphi'_i,\theta,\theta')  \, ,
\end{equation}
where from now on we decide to write all arguments explicitly.
After inserting this in (\ref{ModGreenFunction}) we find
\begin{align}
  &\sum_{\lambda=1}^{\infty}\sum_{l=0}^{\lambda-1} \sum_{m=-l}^l \left[
    -\frac{\lambda^2-1} {R_U^2 \gamma(\theta)^2}c_{\lambda l m}(\varphi'_i,\theta,\theta')+
   \frac{1}{R^2} \frac{1}{\sigma(\theta)\gamma(\theta)^3}\frac{\partial}{\partial\theta}\left(
   \sigma(\theta)\gamma(\theta)^3 
   \frac{\partial c_{\lambda l m}(\varphi'_i,\theta,\theta')}{\partial \theta}\right)\right] 
   \Phi_{\lambda l m}(\varphi_i)\nonumber \\
  &\qquad\qquad \qquad\qquad 
   =\frac{\delta^3(\varphi_i-\varphi'_i) \, \delta(\theta-\theta')}
     {\sqrt{-g}}-u_0(\varphi_i,\theta)\, \overline{u}_0(\varphi'_i,\theta').
\end{align}
After multiplication by $\sqrt{\eta} \, \overline{\Phi}_{\lambda' l' m'}(\varphi_i)$ and integration over 
$S^3$ we obtain
\begin{align}\label{TransverseEquation}
  &-\frac{\lambda'^2-1} {R_U^2 \gamma(\theta)^2}c_{\lambda' l' m'}(\varphi'_i,\theta,\theta')+
   \frac{1}{R^2}\frac{1}{\sigma(\theta)\gamma(\theta)^3} \frac{\partial}{\partial\theta}\left[
   \sigma(\theta)\gamma(\theta)^3 \frac{\partial c_{\lambda' l' m'}(\varphi'_i,\theta,\theta')}
    {\partial \theta}\right] \\ \nonumber
  & \qquad \qquad =\frac{1}{R_U^3 R}\frac{1}{\sigma(\theta)\gamma(\theta)^3} \delta(\theta-\theta')
  \overline{\Phi}_{\lambda' l' m'}(\varphi'_i)-
  \overline{u}_0(\varphi'_i,\theta') \int_{S^3}\sqrt{\eta} \, u_0(\varphi_i,\theta) \, 
  \overline{\Phi}_{\lambda' l' m'}(\varphi_i) \, d^3\varphi_i \, .
\end{align}

We now have to distinguish the cases $\lambda'=1$ and $\lambda'\neq 1$.
\newpage
\begin{enumerate}
  \item $( \lambda' l' m')=(1 0 0)$. 

In this case, the last term on the right hand side of 
(\ref{TransverseEquation}) will give a non-vanishing contribution:
\begin{align}
     &\frac{1}{R^2}\frac{1}{\sigma(\theta)\gamma(\theta)^3} \frac{\partial}{\partial\theta}\left[
     \sigma(\theta)\gamma(\theta)^3 \frac{\partial c_{1 0 0}(\varphi'_i,\theta,\theta')}
      {\partial \theta}\right] =\frac{1}{R_U^3 R} \frac{1}{\sigma(\theta)\gamma(\theta)^3} 
     \delta(\theta-\theta') \overline{\Phi}_{1 0 0}(\varphi'_i)-\\
     &\qquad \vert N_0 \vert^2 \bar{\Phi}_{1 0 0}(\varphi'_i) 
     \chi_1(\theta) \overline{\chi}_1(\theta') \underbrace{\int_{S^3} \sqrt{\eta} \, \Phi_{1 0 0}(\varphi_i) 
     \overline{\Phi}_{1 0 0}(\varphi_i) \, d^3 \varphi_i}_{=1} \nonumber \, .
\end{align}
By defining $g^{(1)}(\theta,\theta')$ by the relation
\begin{equation}
  c_{1 0 0}(\varphi'_i,\theta,\theta')=\frac{1}{R_U^3 R} \bar{\Phi}_{1 0 0}(\varphi'_i) 
   g^{(1)}(\theta,\theta')
\end{equation}
we obtain the following differential equation for $g^{(1)}(\theta,\theta')$:
\begin{equation} \label{TransverseEq1}
  \frac{1}{R^2} \frac{1}{\sigma(\theta)\gamma(\theta)^3} \frac{\partial}{\partial\theta}
    \left[ \sigma(\theta)\gamma(\theta)^3 \frac{\partial g^{(1)}(\theta,\theta')}{\partial \theta}
    \right]=\frac{1}{\sigma(\theta)\gamma(\theta)^3} \delta(\theta-\theta')-
    \tilde{\chi}_1(\theta) \bar{\tilde{\chi}}_1(\theta') \, ,
\end{equation}
where we used 
\begin{equation} 
  \tilde{\chi}_1(\theta)\equiv \left[ \frac{2\omega}{\tanh \left(\frac{\omega\pi}{2}\right)} 
  \right]^{\frac{1}{2}} \chi_1(\theta)\, , \qquad (\chi_1(\theta)\equiv1) \,.
\end{equation}
Note that we defined $\tilde{\chi}_1(\theta)$ in such a way that 
\begin{equation} 
  \int\limits_{-\frac{\pi}{2}}^{\frac{\pi}{2}} \sigma(\theta)\gamma(\theta)^3 \tilde{\chi}_1(\theta)
   \bar{\tilde{\chi}}_1(\theta) d\theta=1.
\end{equation}

\item $(\lambda' l' m')\neq(1 0 0)$. 

Due to the orthogonality of $\Phi_{1 0 0}$ and 
$\Phi_{\lambda' l' m'}$ on $S^3$, the last term on the right hand side of 
(\ref{TransverseEquation}) vanishes. We therefore have
  \begin{align}
     &-\frac{\lambda'^2-1}{R_U^2 \gamma(\theta)^2} c_{\lambda' l' m'}(\varphi'_i,\theta,\theta')+
     \frac{1}{R^2}\frac{1}{\sigma(\theta)\gamma(\theta)^3} \frac{\partial}{\partial\theta}\left[
     \sigma(\theta)\gamma(\theta)^3 \frac{\partial c_{\lambda' l' m'}(\varphi'_i,\theta,\theta')}
      {\partial \theta}\right]\qquad \qquad \nonumber \\
     &\qquad =\frac{1}{R_U^3 R} \frac{1}{\sigma(\theta)\gamma(\theta)^3} \delta(\theta-\theta')
    \bar{\Phi}_{\lambda' l' m'}(\varphi'_i)\, .
  \end{align}
Introducing $g^{(\lambda')}(\theta,\theta')$ again via
\begin{equation}
  c_{\lambda' l' m'}(\varphi'_i,\theta,\theta')=\frac{1}{R_U^3 R} \bar{\Phi}_{\lambda' l' m'}(\varphi'_i)
  g^{(\lambda')}(\theta,\theta')\, ,
\end{equation}
we see that $g^{(\lambda')}(\theta,\theta')$ has to satisfy
\begin{equation} \label{TransverseEq2}
    \frac{1-\lambda'^2}{R_U^2 \gamma(\theta)^2} g^{(\lambda')}(\theta,\theta')+
    \frac{1}{R^2} \frac{1}{\sigma(\theta)\gamma(\theta)^3} \frac{\partial}{\partial\theta}
    \left[ \sigma(\theta)\gamma(\theta)^3 \frac{\partial g^{(\lambda')}(\theta,\theta')}{\partial \theta}
    \right]=\frac{\delta(\theta-\theta')}{\sigma(\theta)\gamma(\theta)^3} \, .
\end{equation}
\end{enumerate}

Combining the above results for $\lambda=1$ and $\lambda\neq 1$ we are able to write the formal solution
 of (\ref{ModGreenFunction}) as
\begin{equation} 
  \mathcal{G}(\varphi_i,\varphi'_i,\theta,\theta')=\frac{1}{R_U^3 R}\sum_{\lambda=1}^{\infty}
  \sum_{l=0}^{\lambda-1}\sum_{m=-l}^l \Phi_{\lambda l m}(\varphi_i)\bar{\Phi}_{\lambda l m}(\varphi'_i)
  g^{(\lambda)}(\theta,\theta') \, .
\end{equation}

We obtain a further simplification of this formal solution by 
employing the spherical symmetry on $S^3$, see eq.~(\ref{SumOverS3}), leaving us with a representation of the 
two-point function by a Fourier sum, which is natural for a compact space without boundaries:
\begin{equation} \label{FormalSolution}
  \tilde{\mathcal{G}}(s,\theta,\theta')=\mathcal{G}(\varphi_i,\varphi'_i,\theta,\theta')
  =\frac{1}{R R_U^3} \sum_{\lambda=1}^{\infty} \frac{\lambda}{2 \pi^2} \frac{\sin(\lambda s)}{\sin s}
  g^{(\lambda)}(\theta,\theta') \, .
\end{equation}

Finding the formal solution (\ref{FormalSolution}) was straightforward apart from minor complications 
inherent to the use of a
modified Green's function. To solve the differential equations (\ref{TransverseEq1}) and 
(\ref{TransverseEq2}) by imposing appropriate boundary conditions again is a routine task without 
any conceptual difficulties, though slightly technical in nature. Replacing back this solution in 
(\ref{FormalSolution}) we are left with a Fourier sum which at first sight looks intractable, 
anticipating the fact that the  solutions $g^{(\lambda)}(\theta,\theta')$ (for $\lambda\neq1$) are
given by hypergeometric functions. Nevertheless it is possible to 
extract the desired asymptotic information from the sum (\ref{FormalSolution}). 

In order not to disturb the transparency and fluidity of the main article we provide large parts 
of the technical calculations in four appendices. 
In appendix \ref{appParallel} we report similarities and 
differences in the evaluation of the two-point functions between our case and the case of Randall-Sundrum, 
since the latter served as a guideline for handling the more difficult case under consideration.
The solutions of eqs. (\ref{TransverseEq1}) and (\ref{TransverseEq2}) are presented in appendix \ref{appDEQ} 
and the evaluation of the Fourier sum (\ref{FormalSolution}) at distances exceeding the size of the extra dimension in 
appendix \ref{appSum}. 
Eventually, appendix \ref{appGreenUltraShort} contains the evaluation of the Fourier sum 
(\ref{FormalSolution}) for distances smaller than the extra dimension.
In this way, we can offer the reader less interested in the details of the computations to have
the main results at hand.

From its definition (\ref{ModGreenFunction}) we understand that the Green's function (\ref{FormalSolution}) 
can be considered as the response of the scalar field to the combination of a point-like source located 
at coordinates $(\varphi'_i, \theta')$ and a delocalized, compensating negative contribution.   
Since we would like to see the response to a point-like particle on the brane we put $\theta'=0$ in eq.
(\ref{FormalSolution}) and explicitly write the $\lambda=1$ term:
\begin{equation} \label{GreenSolBrane}
  \tilde{\mathcal{G}}(s,\theta,0)=\frac{g^{(1)}(\theta,0)}{2 \pi^2 R R_U^3}+
  \frac{R}{4 \pi^2 R_U^3}\frac{1}{\sin s} S[s,\theta,\omega] \, ,
\end{equation}
with $S[s,\theta,\omega]$ given by (\ref{TheSumB}) of appendix \ref{appSum} 
(see also (\ref{glambdaratio}) of appendix \ref{appDEQ}).
The general result for the sum $S[s,0,\omega]$ obtained in appendix \ref{appSum} is 
\begin{align} \label{SresApp}
  S[s,0,\omega] \equiv  \lim_{\theta \to 0} S[s,\theta,\omega] 
  &= -\frac{2}{\omega} \frac{z(0)^\frac{1}{2}}{1-z(0)}
  \left(\frac{\pi-s}{2}\cos s -\frac{1}{4}\sin s \right)\nonumber \\
  &-\frac{1}{2 \omega} z(0)^{-\frac{1}{2}} \ln\left[1-z(0)\right] \sin s
  -\frac{1}{\omega} z(0)^\frac{1}{2} \lim_{\theta \to 0} R[s,\theta,\omega]\;,
\end{align} 
where we have $z(0)= \tanh^2 \left(\frac{\omega\pi}{2} \right)$ and 
where we refer to (\ref{defR}) for the definition of $R[s,\theta,\omega]$. The first term in
(\ref{SresApp}) is the zero mode contribution\footnote{At this point we have to 
explain what we mean by ``zero mode'' in this context, since 
due to the asymmetrically warped geometry, the spectrum of Kaluza-Klein excitations will not 
be Lorentz-invariant and strictly speaking, different excitations cannot be characterized by different 
$4$-dimensional masses. More accurate would be to say that for a given value of the 
momentum eigenvalue $\lambda$ there exists a tower of corresponding Kaluza-Klein excitations 
with energies given by $E_n^\lambda$ with $n=0,1,\ldots$. In our use of language the ``zero mode branch'' 
of the spectrum or simply the ``zero mode'' is defined to be the collection of the lowest energy excitations corresponding to 
all possible values of $\lambda$, 
that is by the set $\left\{\left(E_0^\lambda,\lambda\right),\, \lambda=1,2,\ldots\right\}$. The definition 
is readily generalized to higher branches of the spectrum.} 
and we see that it reproduces exactly the $4$-dimensional 
static Green's function of Einstein's static universe given in (\ref{ESUresult}). 
The other two terms are the contributions from the higher Kaluza-Klein modes.

We first concentrate on the case where $s \sim 1$ or what is equivalent $r\sim R_U$. Since one 
can easily convince oneself that $R[s,0,\omega]\sim \left[1-z(0)\right]^0$ in this regime, the 
contributions of the higher Kaluza-Klein modes are strongly suppressed with respect to the 
zero mode contribution. This means that as in the case of the Randall-Sundrum-II model it 
is the zero mode which dominates the behavior of gravity at distances much larger than the extra
dimensions $R$. The main difference to the Randall-Sundrum-II case is that the zero mode of 
our model not only gives rise to the 
typical $4$-dimensional $1/r$ singularity but also accounts for the compactness of space by
reproducing the Einstein static universe behavior (\ref{ESUresult}). This result is somewhat 
surprising given the extreme anisotropy of our manifold. Due to the fact that the distances 
between two arbitrary points on the brane are of the order of $R$, one might intuitively 
expect that the extra dimension can be effective in determining also the large distance 
behavior of gravity (on the brane). As we could show by direct calculation the above expectation 
turns out to be incorrect. 

Next we consider physical distances $r$ much larger than the extra dimension $r\gg R$ and much smaller 
than the observable universe $r \ll R_U$
in which case the results for $R[s,0,\omega]$ can be seen to be:
\begin{align} 
  R[s,0,\omega]&\sim\frac{\pi}{2 s^2}+
  \frac{\pi}{2}\frac{1-z(0)}{s^4} \left\{ 8-6\ln 2 -6 \ln\left[s \left[1-z(0)\right]^{-1/2}\right]\right\} 
  \nonumber \\
  &+\mathcal{O}\left[\left[1-z(0)\right]^2 \frac{\ln\left[s \left[1-z(0)\right]^{-1/2}\right]}{s^6}\right]
\end{align}
valid for $\left[1-z(0)\right]^{1/2} \ll s \ll 1$.
The zero mode contribution (to $S[s,0,\omega]$) in this regime is simply the constant obtained by 
setting $s=0$ in the first term of (\ref{SresApp}) so that we obtain:
\begin{equation} 
  S[s,0,\omega] \sim -\frac{\pi}{\omega} \frac{z(0)^{\frac{1}{2}}}{1-z(0)} 
  \left\{1+\frac{1}{2 \bar{s}^2}+\frac{1}{\bar{s}^4}\left[4-3 \ln 2 - 3 \ln \bar{s} \right] +
  \mathcal{O} \left( \frac{\ln \bar{s}}{\bar{s}^6}\right)\right\} \, ,
\end{equation}
where we introduced $\bar{s}=s \left[1-z(0)\right]^{-1/2}$. Inserting this result in (\ref{GreenSolBrane})
and using physical distance $r=R_U s$ instead of $s$ we obtain
\begin{align} \label{FullCorrections}
  \tilde{\mathcal{G}}(s,0,0)&=\frac{g^{(1)}(0,0)}{2 \pi^2 R R_U^3}-
  \frac{1}{4 \pi r} \frac{\omega}{R} \coth\left( \frac{\omega \pi}{2}\right)
 \left\{1+\frac{1}{2 \bar{s}^2}+\frac{1}{\bar{s}^4}\left[4-3 \ln 2 - 3 \ln \bar{s} \right] +
  \mathcal{O} \left( \frac{\ln \bar{s}}{\bar{s}^6}\right)\right\}=\nonumber \\
  &=\frac{g^{(1)}(0,0)}{2 \pi^2 R R_U^3}-
  \frac{1}{4 \pi r} \frac{\omega}{R} \coth\left( \frac{\omega \pi}{2}\right)
  \left\{1+\frac{\tanh^2\left(\frac{\omega\pi}{2}\right)}{2 \bar{r}^2}+
   \frac{\tanh^4\left(\frac{\omega\pi}{2}\right)}{\bar{r}^4} \times \right. \nonumber \\ 
  &\quad \times \left. \left[4-3 \ln 2 
  - 3 \ln \left(\frac{\bar{r}}{\tanh \left(\frac{\omega \pi}{2}\right)}\right) \right]
  + \mathcal{O} \left[ \frac{\tanh^6\left(\frac{\omega\pi}{2}\right)}{\bar{r}^6} 
    \ln \left(\frac{\bar{r}}{\tanh \left(\frac{\omega \pi}{2}\right)}\right) \right]\right\} \, .
\end{align}
Since we would like to compare our result with the corresponding correction in the Randall-Sundrum II
case, we introduced the dimensionless distance variable $\bar{r}=r \omega/R$, 
the physical distance measured in units of the AdS-radius, in the last line of the above result. 
We see that 
in complete agreement with the Randall-Sundrum II scenario, our setup reproduces 
$4$-dimensional gravity at large distances with extremely suppressed corrections. The only remnant 
effect from the different global topology manifest itself through the factors of 
$\tanh \left(\frac{\omega \pi}{2}\right)$ and $\coth \left(\frac{\omega \pi}{2}\right)$ which 
are very close to $1$. 
We furthermore emphasize that
apart from these deviations, the asymptotic we obtained coincides exactly with the 
asymptotic for the case of a massless scalar field in 
the Randall-Sundrum-II background, see (\ref{I1Asymptotic})
and e.g.~\cite{Callin:2004py,Kiritsis:2002ca,Ghoroku:2003bs,Giddings:2000mu}. 
From the factor $\omega \coth \left(\frac{\omega \pi}{2}\right)/R$ 
in eq. (\ref{FullCorrections}) we see that also the relation between the fundamental scale $M$ and 
the Planck-scale $M_{Pl}$ gets modified only by the same factor of 
$\tanh\left(\frac{\omega \pi}{2}\right)$:
\begin{equation} 
  M_{Pl}^2 = M^3 \frac{R}{\omega} \tanh \left(\frac{\omega \pi}{2} \right) \, ,
\end{equation}
where we remind that $R/\omega$ is nothing but the AdS-Radius.

Eventually, we treat the case of distances inferior to the extra dimension $r\ll R$. 
After using the result (\ref{5DNewton}) of appendix \ref{appGreenUltraShort} in (\ref{GreenSolBrane}), a 
short calculation reveals 
\begin{align} \label{Short5DCorrections}
  \tilde{\mathcal{G}}(s,0,0)&=\frac{g^{(1)}(0,0)}{2 \pi^2 R R_U^3}-\frac{1}{4\pi^2}
  \frac{1}{R_U^2 s^2+R^2 \theta^2}, 
\end{align}
a result that has to be compared to the characteristic solution of the Poisson equation in 4-dimensional
flat space. In $n$-dimensional flat space one has:
\begin{equation} \label{CharLaplSolnDspace}
  \Delta \left[ -\frac{1}{(n-2) V_{S^{n-1}} r^{n-2}}\right]=\delta(r) \, , \qquad 
  r=\left(\sum_{i=1}^n x_i^2\right)^{\frac{1}{2}}
\end{equation}
with $V_{S^{n-1}}=2\pi^{\frac{n}{2}}/\Gamma[\frac{n}{2}]$ denoting the volume of the $n-1$ sphere. Specifying
to $n=4$, we recover the correct prefactor of $-1/4\pi^2$ in (\ref{Short5DCorrections}) multiplying the 
$1/r^2$ singularity.

Finally, we mention that in none of the considered cases we payed any attention to the additive constant 
in the two-point function on the brane. The arbitrary constant entering the solution 
$g^{(1)}(\theta,\theta')$ can always be chosen in such a way that $g^{(1)}(0,0)$ vanishes on the brane (see
appendix~\ref{appDEQ}).

\section{Conclusions}
\label{four}

In this paper we considered a particular brane world model in 5 dimensions with the 
characteristic property that the spatial part of the space-time manifold (including the 
extra dimension) is compact and has the topology of a $4$-sphere $S^4$. Similar to the 
original Randall-Sundrum~II model, the $3$-brane is located at the boundary between two
regions of $AdS_5$ space-time. The coordinates of $AdS_5$ used by Randall and Sundrum 
are closely related to the so-called Poincar\'e coordinates of $AdS_5$. While the extra 
dimension in this set of coordinates provides a slicing of $AdS_5$ along flat $4$-dimensional
Minkowski sections (resulting in a flat Minkowskian induced metric on the brane), 
their disadvantage is that they do not cover the whole of $AdS_5$ space-time. 

The coordinates we used in this paper are the global coordinates of $AdS_5$ known to provide 
a global cover of the $AdS_5$ space-time. In this case the ``extra'' dimension labels different
sections with intrinsic geometry $\mathbb{R} \times S^3$, the geometry of 
Einsteins static universe. The induced metric on the $3$-brane in our setup is therefore also 
given by $\mathbb{R} \times S^3$.

As we illustrated with the use of the Penrose-diagram of $AdS_5$, the incompleteness of the 
Poincar\'e patch is at the origin of the incompleteness of the Randall Sundrum II 
space-time with respect to timelike and lightlike geodesics. Moreover we were able to demonstrate 
that the setup considered in this paper provides an alternative to the Randall-Sundrum II model 
which does not suffer from the drawback of being geodesically incomplete. The latter point was 
part of the main motivations for this work.

The spatial 
part of our manifold is characterized by an extreme anisotropy with respect to one of the 
coordinates (the extra coordinate) accounting for thirty orders of magnitude between the
size of the observable universe and present upper bounds for the size of extra dimensions. 

Another interesting property of our manifold related to the anisotropy of its spatial part is the
fact that {\em any two points} on the brane are separated by a distance of the order of the 
size of the extra dimension $R$ regardless of their distance measured by means of the 
induced metric on the brane. Despite the difference in the global topology, the properties of 
gravity localization turned out to be very similar to the Randall-Sundrum~II model, though much more 
difficult to work out technically. We computed the static (modified) Green's function of a massless 
scalar field in our background and could show that in the intermediate distance regime 
$R \ll r \ll R_U$ the $4$-dimensional Newton's law is valid for two test particles on the brane, 
with asymptotic corrections terms identical to the Randall-Sundrum~II case up to tiny 
factors of $\tanh\left(\frac{\omega\pi}{2}\right)$. We could also 
recover the characteristic $5$-dimensional behavior of the Green's function for distances smaller than
the extra dimension $r \ll R$. Eventually we saw that in the regime of cosmic distances 
$r \sim R_U$, somewhat 
counterintuitive given our highly anisotropic manifold, the Green's function is dominated 
by the behavior of the corresponding static (modified)
Green's function in Einstein's static universe.

In the simple setup considered in this paper the $3$-brane is supposed to be motionless.
In the light of recent progress in the study of $4$-dimensional cosmic evolution
induced by the motion of the brane in the bulk, it would be interesting to explore this possibility and 
see what kind of modifications of our results we would have to envisage. Finally a related, 
important question which would be interesting to address would be the question of stability of our setup.

\vspace{1cm}
\noindent {\it Acknowledgments:} We wish to thank S.~Dubovsky, P.~Tinyakov and S.~Khlebnikov 
for useful comments and discussions. 
E.~R. is particularly grateful to E.~Teufl for helpful advice in numerous mathematical questions. 
A.~G. wishes to thank LPPC for the kind hospitality during most of this
research. This work was supported by the Swiss Science Foundation. 
A.~G. acknowledges "Fondazione A. Della Riccia" for financial support. 

\newpage
\appendix

\section{Parallels in the computation of the corrections to Newton's law between the case under 
consideration and the Randall-Sundrum II case} \label{appParallel}
\setcounter{equation}{0}
The purpose of this appendix is to review briefly the calculations of the static two-point function of 
a scalar field in the Randall-Sundrum II background \cite{Kiritsis:2002ca,Giddings:2000mu} and to 
compare each stage with the corresponding 
stage of calculations in the background considered in this paper. This serves mainly for underlining
similarities and differences between the two calculations. Let us begin by writing down the metric of 
the Randall-Sundrum II model
\begin{equation} \label{RSmetric}
  ds^2=e^{-2 k \vert y\vert} \eta_{\mu \nu} dx^\mu dx^\nu + dy^2\, , \qquad -\pi r_c \leq y \leq \pi r_c \, .
\end{equation}
Here $y$ stands for the extra dimension while $k$ and $\eta_{\mu \nu}$ denote the inverse radius of $AdS_5$
space and the (4-dimensional)-Minkowski metric with signature $-+++$. Due to the orbifold $Z_2$-symmetry the
allowed range of $y$ is $0 \leq y \leq \pi r_c$.\footnote{We allow for a finite $r_c$ only to impose 
boundary conditions in a proper way. Eventually we are interested in the limit $r_c \to \infty$.}
As it is well known, each $4$-dimensional graviton mode in this background satisfies the
 equation of a massless scalar field. 
Therefore, for the study of the potential between two test masses on the brane we confine ourselves 
to solving the equation of a massless scalar field with an arbitrary time-independent source:
\begin{equation} \label{RSScalarEq}
  \mathcal{D} u(\vec x,y) = j(\vec x,y) \qquad \mbox{with} \; \; \mathcal{D}=e^{2 k y} \Delta_x -
   4 k\frac{\partial}{\partial y}+\frac{\partial^2}{\partial y^2} \, .
\end{equation}
Since the operator $\mathcal{D}$ is formally self-adjoint, the corresponding Green's function
will satisfy:
\begin{equation} \label{RSGreendef}
  \mathcal{D} \mathcal{G}(\vec x,\vec x';y,y')=\frac{\delta^3(\vec x-\vec x') \, \delta(y-y')}{\sqrt{-g}} \, .
\end{equation}
We are now able to write down the usual integral representation of the solution of (\ref{RSScalarEq}):
\begin{equation} \label{RSSolIntRep}
  u(\vec x,y)=\int \sqrt{-g} \, j(\vec x',y')\,  \overline{G(\vec x,\vec x';y,y')} \, d^3 \vec x' dy',
\end{equation}
where the $y'$-integration extends from $0$ to $\pi r_c$ and the $\vec x'$ integrations from
$-\infty$ to $\infty$.
The absence of boundary terms in (\ref{RSSolIntRep}) is of course the result of an appropriate choice 
of  boundary conditions for $u(\vec x,y)$ and  $G(\vec x,\vec x';y,y')$. In the above coordinates 
of the Randall-Sundrum II case the orbifold boundary conditions together with the Israel condition 
imposed on the fluctuations of the metric give rise to a Neumann boundary condition at $y=0$. One can easily 
convince oneself that in the limit $r_c \to \infty$ the resulting Green's 
function is independent of the choice of the (homogeneous) boundary condition at $y=\pi r_c$. We therefore
follow \cite{Randall:1999vf} and use also a Neumann boundary condition at $y=\pi r_c$. Another important 
point is that the Green's function we are considering is specific to the orbifold boundary condition and
so describes the situation of a semi-infinite extra dimension.
We decided to carry out the calculations in the semi-infinite case as opposed to \cite{Randall:1999vf}, 
where eventually the orbifold boundary conditions are dropped and the case of a fully infinite extra 
dimension is considered.\footnote{However as long as matter on the 
brane is considered the Green's function for the two setups differ only by a factor of $2$.}
An immediate consequence of this will be that the constant factors of the 
characteristic short distance singularities of the solutions of Laplace 
equations also will be modified by a factor of $2$. Finally we note that this factor of $2$ can be 
accounted for by an overall redefinition of the 5-dimensional Newton's constant leaving the two theories
with equivalent predictions. We conclude the discussion of boundary conditions by noting that
at infinity in $\vec x'$ we suppose that the $u(\vec x',y')$ and $G(\vec x, \vec x'; y,y')$ vanish 
sufficiently rapidly.

The solution of eq.~(\ref{RSGreendef}) can be found most conveniently by Fourier expansion 
in the $\vec x$-coordinates and by direct solution of the resulting differential equation for the 
 ``transverse'' Green's function. After performing the integrations over the angular coordinates in
Fourier space we obtain:
\begin{equation} \label{RSGreenFormalSolution}
  \mathcal{G}(\vec x-\vec x',y,y')=\frac{1}{2 \pi^2}\int\limits_0^\infty 
  \frac{p \sin\left(p \vert \vec x-\vec x' \vert \right)}{\vert \vec x-\vec x' \vert} g^{(p)}(y,y') dp \, ,
\end{equation}
where $g^{(p)}(y,y')$ satisfies
\begin{equation} \label{RSTransGreen}
  \frac{1}{e^{-4 k y}} \frac{\partial}{\partial y}
   \left[e^{-4 k y} \frac{\partial}{\partial y} g^{(p)}(y,y')\right]-p^2 e^{2 k y} g^{(p)}(y,y')=
  \frac{\delta(y-y')}{e^{-4 k y}}\, , \qquad (0 \leq y \leq \pi r_c)
\end{equation}
together with Neumann boundary conditions at $y=0$ and $y=\pi r_c$.
The solution of (\ref{RSTransGreen}) is straightforward with the general result
\begin{align}
  & g^{(p)}(y,y')=\frac{e^{2 k \left(y_>+y_< \right)}}{k} \times \\
  & \;\; \times \frac{\left[I_1\left(\frac{p}{k}\right) K_2\left(\frac{p}{k} e^{k y_<}\right)
        +K_1\left(\frac{p}{k}\right) I_2\left(\frac{p}{k} e^{k y_<}\right) \right] \cdot
        \left[I_1\left(\frac{p}{k} e^{k \pi r_c} \right) K_2\left(\frac{p}{k} e^{k y_>}\right)
        +K_1\left(\frac{p}{k} e^{k \pi r_c} \right) I_2\left(\frac{p}{k} e^{k y_>}\right) \right]}
    {I_1\left( \frac{p}{k}\right) K_1\left( \frac{p}{k} e^{k \pi r_c} \right)-
     I_1\left( \frac{p}{k} e^{k \pi r_c} \right) K_1\left( \frac{p}{k}\right)} \nonumber \, ,
\end{align}
where $y_>$ ($y_<$) denote the greater (smaller) of the two numbers $y$ and $y'$ and 
$I_1(z), I_2(z), K_1(z), K_2(z)$ are modified Bessel functions.
Since we want to study sources on the brane we now set $y'=0$ and take the well-defined 
limit $r_c\to \infty$, as can easily be verified from the asymptotic behavior of $I_n$ and $K_n$
for large values of $z$. We are therefore able to write the static scalar two-point function 
as
\begin{eqnarray} \label{RSGreenSol}
  \mathcal{G}(\vec x-\vec x';y,0)=-\frac{e^{2 k y}}{2 \pi^2} \int\limits_0^\infty
  \frac{\sin\left(p \vert \vec{x}-\vec x' \vert \right)}{\vert \vec x-\vec x' \vert}
  \frac{K_2 \left(\frac{p}{k} e^{k y} \right)}{K_1 \left(\frac{p}{k} \right)} \, dp \, .
\end{eqnarray}
We would like to point out a particularity of the Fourier-integral (\ref{RSGreenSol}). By using the 
asymptotic expansions for the modified Green's functions $K_1(z)$ and $K_2(z)$ we have \cite{AbrStegun}:
\begin{equation} \label{KoverKAsymptotic}
  \frac{K_2 \left(\frac{p}{k} e^{k y} \right)}{K_1 \left(\frac{p}{k} \right)} \sim 
  e^{-\frac{1}{2} k y} e^{-\frac{p}{k}\left(e^{k y}-1\right)}\left[1+\mathcal{O}\left(\frac{1}{p}
  \right)\right],
\end{equation}
showing that for $y=0$ the integral over $p$ in (\ref{RSGreenSol}) does not exist. The correct 
value of the Green's function on the brane is therefore obtained by imposing continuity at $y=0$:
\begin{equation} 
  \mathcal{G}(\vec{x}-\vec{x}';0,0)\equiv\lim_{y \to 0} \mathcal{G}(\vec{x}-\vec{x}';y,0).
\end{equation}

Note that this subtlety is still present in our case of the Fourier sum representation of the modified
 Green's function (\ref{FormalSolution}), as discussed in appendix \ref{appSum}. 

The representation (\ref{RSGreenSol}) is very suitable for obtaining the short distance behavior of the
Green's function. By inserting the expansion (\ref{KoverKAsymptotic}) into (\ref{RSGreenSol}) we can evaluate the 
integral which will give reasonable results for distances smaller than $1/k$:
\begin{equation} 
  \mathcal{G}(\vec x-\vec x';y,0)\sim-\frac{e^{\frac{3}{2} k y}}{2 \pi^2}
  \frac{1}{\vert \vec x-\vec x' \vert^2+y^2} \, .
\end{equation}
We see that after replacing the exponential factor by $1$ (valid for $y \ll 1/k$) we 
recover (up to a factor of $2$) the correct $5$-dimensional behavior of the Green's function in flat 
4-dimensional space (\ref{CharLaplSolnDspace}).\footnote{As discussed above the factor of $2$ is 
a direct consequence of the boundary conditions corresponding to a semi-infinite extra dimension.}

For large distances the Fourier-representation (\ref{RSGreenSol}) is less suited for obtaining corrections 
to Newton's law. This is due to the fact that all but the first term in the expansion of 
$K_2/K_1$ in powers of $p$ around $p=0$ lead to divergent contributions upon inserting 
in (\ref{RSGreenSol}). We therefore seek another method which will allow us to obtain the corrections 
to Newton's law by term-wise integration. 

\begin{figure}[htbp]
\begin{center}
\input{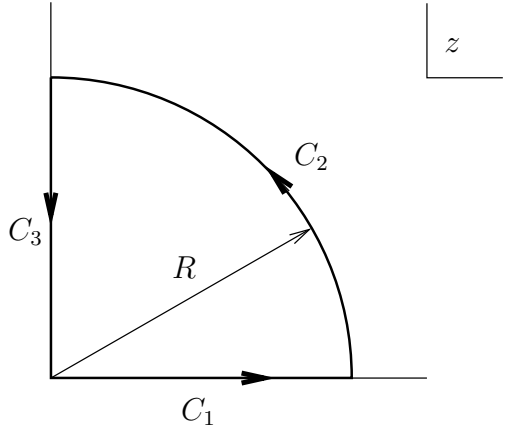}
\end{center}
\caption{Contour used in \textsc{Cauchy}'s theorem to evaluate the integral (\ref{RSI1FourierRep}).}
\label{intcontour}
\end{figure}

The idea is to promote (\ref{RSGreenSol}) to a contour-integral in the complex $p$-plane and to
shift the contour in such a way that the trigonometric function is transformed into 
an exponential function. First we introduce dimensionless quantities by rescaling with $k$ according to
$X=\vert \vec x - \vec x' \vert k$, $Y = y k$, $z=p \vert \vec x - \vec x' \vert$:
\begin{equation} 
  \mathcal{\bar{G}}(X,Y,0)=-\frac{k^2 e^{2 Y}}{2 \pi^2 X^2}
  \underbrace{\int\limits_0^\infty \sin z \frac{K_2\left(\frac{z}{X} e^Y \right)}{K_1\left(\frac{z}{X} 
   \right)} dz}_{\equiv I\left[X,Y\right]} \, .
\end{equation}
Following \cite{Kiritsis:2002ca} (see also \cite{AbrStegun}) we now use the relation 
\begin{equation} \label{BesselKRelation}
  K_2[w]=K_0[w]+\frac{2}{w}K_1[w] 
\end{equation}
to separate the zero mode contribution to the Green's function (Newton's law) from the contributions 
coming from the higher Kaluza-Klein particles (corrections to Newton's law).
Using (\ref{BesselKRelation}) in $I[X,Y]$ we find 
\begin{equation} 
  I[X,Y]=\underbrace{\int\limits_0^\infty \sin z \frac{K_0\left(\frac{z}{X} e^Y \right)}
    {K_1\left(\frac{z}{X}\right)} dz}_{\equiv I_1[X,Y]}+
    \underbrace{\frac{2 X}{e^Y} \int\limits_0^\infty \frac{\sin z}{z} 
   \frac{K_1\left(\frac{z}{X} e^Y \right)}{K_1\left(\frac{z}{X} \right)} dz}_{\equiv I_2[X,Y]} \, .
\end{equation}
The additional factor of $1/z$ in the integrand of $I_2[X,Y]$ allows us to take the limit $Y \to 0$ with the 
result:
\begin{equation} 
  \lim_{Y \to 0} I_2[X,Y]=\pi X \, .
\end{equation} 
However, we still need $Y>0$ for convergence in $I_1[X,Y]$. The next step is to use 
$\sin z=\Im \left\{e^{\imath z}\right\}$ and to exchange the operation $\Im$ with the integration 
over z:\footnote{This is justified since both the real and the imaginary part of the 
resulting integral converge.}
\begin{equation} \label{RSI1FourierRep}
  I_1[X,Y]=\Im\left\{\int\limits_0^\infty e^{\imath z} 
  \frac{K_0\left(\frac{z}{X} e^Y \right)}{K_1\left(\frac{z}{X} \right)} dz\right\} \, .
\end{equation}
The integrand in (\ref{RSI1FourierRep}) is a holomorphic function of $z$ in the first 
quadrant (see \cite{AbrStegun}, p.377 for details). We can therefore apply \textsc{Cauchy}'s 
theorem to the contour depicted in Fig.~\ref{intcontour}.
As we will show shortly, the contribution from the arc
$C_2=\left\{z \, |\,  z=R e^{\imath \varphi},\;0\leq \varphi \leq \frac{\pi}{2} \right\}$ 
vanishes in the limit $R \to \infty$ (the limit we are interested in).
Symbolically we therefore have
\begin{equation} 
  \lim_{R \to \infty} \left\{\int_{C_1} \ldots + \int_{C_3} \ldots  \right\}=0\, ,
\end{equation}
where $\ldots$ replace the integrand in (\ref{RSI1FourierRep}) and the sense of integration is as
indicated in the figure. This means that we can 
replace the integration over the positive real axis in (\ref{RSI1FourierRep}) by an integration 
over the positive imaginary axis without changing the value of $I_1[X,Y]$. Substituting now $z$ in
favor of $n$  according to $n=-\imath z$ we obtain:
\begin{align} \label{I1closetoresult}
  I_1[X,Y]&=\Im \left\{ \int\limits_0^\infty e^{-n} 
  \frac{K_0\left(\frac{\imath n}{X} e^Y \right)}{K_1\left(\frac{\imath n}{X} \right)} \imath dn \right\}=
  \Im \left\{ \int\limits_0^\infty e^{-n} 
  \frac{-H_0^{(2)} \left(\frac{n}{X} e^Y \right)}{H_1^{(2)}\left(\frac{n}{X} \right)} dn \right\}\nonumber \\
  &=\int\limits_0^\infty e^{-n} \frac{Y_0\left(\frac{n}{X} e^Y\right) J_1\left(\frac{n}{X}\right)-
                               J_0\left(\frac{n}{X} e^Y\right) Y_1\left(\frac{n}{X}\right) }
                               {\left[J_1\left(\frac{n}{X} \right)\right]^2 +
                                \left[Y_1\left(\frac{n}{X} \right)\right]^2} \, dn \, .
\end{align}
In the first line in (\ref{I1closetoresult}) we replaced modified Bessel functions with imaginary arguments
by Hankel functions of real argument. In the second line we explicitly took the imaginary part after replacing
the identities relating the Hankel functions and the Bessel functions $J$ and $Y$ of the first kind. 
The benefit from the rotation of the contour of integration is obvious at this stage. First, the last 
integral in (\ref{I1closetoresult}) converges also for $Y=0$ due to the presence of the exponential
function in the integrand. In this case we have
\begin{equation} \label{RSMainResult}
  I_1[X,0]=\frac{2 X}{\pi} \int\limits_0^\infty \frac{e^{-n}}{n}
  \frac{1}{\left[J_1\left(\frac{n}{X} \right)\right]^2 + \left[Y_1\left(\frac{n}{X} \right)\right]^2} 
  \, dn \, .
\end{equation}
Note that we used the Wronskian relation for Bessel functions to simplify the numerator, the general
 reference being once more \cite{AbrStegun}. 
Second, it is straightforward to obtain the large $X$ asymptotic ($1\ll X $) of the integral 
(\ref{RSMainResult}) by a simple power series expansion in $n$ around $n=0$ of the integrand 
(apart from the exponential factor), followed by term-wise integration. For the results see 
\cite{Callin:2004py}, 
where the above integral (\ref{RSMainResult}) has been found in the wave-function approach:
\begin{equation} \label{I1Asymptotic}
  I_1[X,0]\sim\frac{\pi}{2 X}+\frac{\pi}{X^3}\left(4-3 \ln 2 - 3\ln X \right) + 
  \mathcal{O}\left[\frac{\ln X}{X^5}\right].
\end{equation}
The result for $\mathcal{\bar{G}}(X,0,0)$ is therefore:
\begin{equation} 
  \mathcal{\bar{G}}(X,0,0)=-\frac{k^2}{2 \pi X}
  \left[1+\frac{1}{2 X^2}+\frac{1}{X^4} \left(4-3 \ln 2 - 3\ln X \right) + 
  \mathcal{O}\left(\frac{\ln X}{X^6}\right)\right].
\end{equation} 

However, we still have to verify that the contribution from the arc $C_2$ in 
Fig.~\ref{intcontour} vanishes in the limit $R \to \infty$. Parametrizing $z$ by 
$z=R e^{\imath \varphi}, \; 0 \leq \varphi \leq \pi/2$ we find:
\begin{align} \label{KoverKestimate}
  \left| \int_{C_2} e^{\imath z} 
  \frac{K_0\left(\frac{z}{X} e^Y \right)}{K_1\left(\frac{z}{X} \right)} dz \right|
  &= \left| \int\limits_0^{\frac{\pi}{2}} e^{\imath R e^{\imath \varphi}} 
     \frac{K_0\left(\frac{R e^{\imath \varphi}}{X} e^Y \right)}
      {K_1\left(\frac{R e^{\imath \varphi}}{X} \right)} \left( \imath R e^{\imath \varphi} \right) 
      d\varphi \right| \nonumber \\
  &\leq R \int\limits_0^\frac{\pi}{2} \left| e^{\imath R \left(\cos \varphi+ \imath \sin \varphi\right)}\right|
      \cdot \left| \frac{K_0\left(\frac{R e^{\imath \varphi}}{X} e^Y \right)}
      {K_1\left(\frac{R e^{\imath \varphi}}{X} \right)}\right|  d\varphi \, .
\end{align}
Making now use of the asymptotic properties of the modified Bessel functions to estimate the second modulus 
in the last integrand in (\ref{KoverKestimate}) we can write:
\begin{equation} 
  \left| \frac{K_0\left(\frac{R e^{\imath \varphi}}{X} e^Y \right)}
      {K_1\left(\frac{R e^{\imath \varphi}}{X} \right)}\right| 
   \leq C \left| e^{-\frac{Y}{2}} e^{-R e^{\imath \varphi}\frac{e^Y-1}{X}}\right| \, ,
\end{equation}
so that
\begin{align} 
  \left| \int_{C_2} e^{\imath z} 
  \frac{K_0\left(\frac{z}{X} e^Y \right)}{K_1\left(\frac{z}{X} \right)} dz \right| <
  C R e^{-\frac{Y}{2}} \int\limits_0^{\frac{\pi}{2}} e^{-R\left[ \sin\varphi + 
  \left( \frac{e^Y-1}{X} \right) \cos \varphi \right]} d\varphi \, ,
\end{align}
with $C$ being a constant of the order of unity, independent of $R$ and $\varphi$.
To finish the estimate we observe that 
\begin{equation} 
  \sin \varphi + \left( \frac{e^Y-1}{X} \right) \cos \varphi \geq \epsilon
  \equiv \min \left[1, \frac{e^Y-1}{X} \right] > 0 \quad , \quad \forall\, \varphi \in [0,\frac{\pi}{2}]\, ,
  \qquad (Y>0)
\end{equation}
and obtain:
\begin{equation} \label{arcresult}
  \left| \int_{C_2} e^{\imath z} 
  \frac{K_0\left(\frac{z}{X} e^Y \right)}{K_1\left(\frac{z}{X} \right)} dz \right| <
  \frac{C R \pi}{2} e^{-\frac{Y}{2}}  e^{-R \epsilon} \; \to 0 \quad \mbox{for} \; R \to \infty.
\end{equation}
This establishes the result that in the limit $R \to \infty$ the arc $C_2$ does not contribute to the 
integral in (\ref{RSI1FourierRep}) and completes our discussion of the two-point function in the 
Randall Sundrum-II setup. 

We are now going to summarize briefly the complications arising when 
the general scheme of computations outlined above for the Randall-Sundrum II case are applied to the brane
setup considered in this paper. First and foremost, the main difference is due to the fact that the 
induced metric on our brane is not the flat Minkowski metric but the metric of the Einstein static universe.
This implies the use of a modified Green's function. In the eigenfunction expansion, the discrete 
scalar-harmonics on $S^3$ will replace the continuous plane-wave eigenfunctions. 
We therefore expect the formal solution for the two-point function (the analog of 
(\ref{RSGreenFormalSolution})) to take the form of a Fourier sum.

The solution to the \textsl{transverse} Green's function (the analog of eq. (\ref{RSTransGreen})) as 
presented in appendix \ref{appDEQ} turns out to be again straightforward and completes the formal solution. 

When trying to distinguish between the zero mode contribution and the contribution coming from 
higher modes (in the language of the Kaluza-Klein approach) a relation similar to (\ref{BesselKRelation}) 
proves useful. While the evaluation of the zero mode contribution poses no problems, the corrections 
coming from higher modes are much more involved this time. We face the following 
major technical difficulties: there is no analog of \textsc{Cauchy}'s 
theorem in the discrete case of the Fourier sum. One solution to this problem is to use a variant of 
\textsc{Euler-Maclaurin}'s sum rule in order to replace the Fourier sum by an analog Fourier integral and  
a remainder term also in the form of an integral. Now the procedure again is similar to the one outlined for 
the Randall-Sundrum case. Another technical problem, however, in connection with large distances 
corrections has its origin in the simple fact that by large distances in our setup 
we mean distances large with respect to the extra dimension $R$ but also small with respect to 
the size of the observable universe $R_U$.
The detailed computations of the corrections in this distances regime can be found 
in appendix \ref{appSum}.

What concerns the computation of the two-point function at distances smaller than the extra dimension, there
is no conceptual difference to the Randall-Sundrum case. The Fourier sum can be calculated 
analytically after inserting the corresponding large momentum asymptotic.  We give the details in appendix 
\ref{appGreenUltraShort}.

\newpage
\section{Solutions to the differential equations (\ref{TransverseEq1}) and (\ref{TransverseEq2})} 
\label{appDEQ}
\setcounter{equation}{0}
The aim of this appendix is to obtain the solutions to the eqs.~(\ref{TransverseEq1}) and 
(\ref{TransverseEq2}) and 
hence to complete the computation of the modified Green's function. 
We will concentrate mainly on (\ref{TransverseEq2}) for two reasons: first and foremost because
in the formal expansion (\ref{FormalSolution}) $g^{(1)}(\theta,\theta')$ multiplies a constant function
in $s$ and since we are mainly interested in the behavior of the Green's function on the brane, this constant
is of no relevance for our considerations. Secondly because the construction of the Green's functions 
$g^{(1)}(\theta,\theta')$ and $g^{(\lambda)}(\theta,\theta')$ follows the standard procedure, so it is not 
necessary to go into details twice. We will merely state the result for $g^{(1)}(\theta,\theta')$. 
The equation we would like to solve is 
\begin{equation} \label{TransvGreen}
    \frac{1-\lambda^2}{R_U^2 \gamma(\theta)^2} g^{(\lambda)}(\theta,\theta')+
    \frac{1}{R^2} \frac{1}{\sigma(\theta)\gamma(\theta)^3} \frac{\partial}{\partial\theta}
    \left[ \sigma(\theta)\gamma(\theta)^3 \frac{\partial g^{(\lambda)}(\theta,\theta')}{\partial \theta}
    \right]=\frac{\delta(\theta-\theta')}{\sigma(\theta)\gamma(\theta)^3} \, ,
\end{equation}
where $\lambda=2,3,4\ldots$ and the independent variables $\theta$ and $\theta'$ are restricted to the 
intervals $-\pi/2\leq\theta\leq\pi/2$, 
and $-\pi/2\leq\theta'\leq\pi/2$. 
The homogeneous equation associated with (\ref{TransvGreen}) can be reduced to the hypergeometric 
equation. Its general formal solution in the interval $0\leq\theta\leq\pi/2$ is given by
\begin{eqnarray} \label{TransvGreenGenSol}
  \varphi^{(+)}(\lambda,\theta)=a_1^{(+)} \varphi_1(\lambda,\theta)+a_2^{(+)} \varphi_2(\lambda,\theta) 
  \, , \qquad   0\leq\theta\leq\frac{\pi}{2}
\end{eqnarray}
where 
\begin{align} 
  \varphi_1(\lambda,\theta)&=\frac{\tanh^{\lambda-1}\left[\omega\left(\frac{\pi}{2}-\theta\right)\right]}
  {\cosh^4\left[\omega\left(\frac{\pi}{2}-\theta\right)\right]}
  \phantom{1}_2 F_1\left[\frac{\lambda+3}{2},\frac{\lambda+3}{2};\lambda+1;
        \tanh^2\left[\omega\left(\frac{\pi}{2}-\theta\right)\right]\right] \, ,\nonumber \\ 
  \varphi_2(\lambda,\theta)&=\frac{\tanh^{\lambda-1}\left[\omega\left(\frac{\pi}{2}-\theta\right)\right]}
  {\cosh^4\left[\omega\left(\frac{\pi}{2}-\theta\right)\right]}
  \phantom{1}_2 F_1\left[\frac{\lambda+3}{2},\frac{\lambda+3}{2};3;
        \cosh^{-2}\left[\omega\left(\frac{\pi}{2}-\theta\right)\right]\right] \, .
\end{align}
Due to the symmetry of the background under $\theta\to -\theta$ we can immediately write down the 
general solution of the homogeneous equation obtained from (\ref{TransvGreen}) in the interval
$-\pi/2\leq\theta\leq 0$:
\begin{eqnarray} \label{TransvGreenGenSolNeg}
  \varphi^{(-)}(\lambda,\theta)=a_1^{(-)} \varphi_1(\lambda,-\theta)+a_2^{(-)} \varphi_2(\lambda,-\theta) \, ,
  \qquad -\frac{\pi}{2}\leq\theta\leq 0 \, .
\end{eqnarray}
We now have to decide whether we want to restrict our Green's function to describe perturbations 
which also possess the symmetry $\theta\to-\theta$ of the background (as in the Randall-Sundrum case) or not. 
In the first case one can limit the solution of (\ref{TransvGreen}) to $\theta>0 $ and continue the 
result symmetrically into $\theta<0$. Note that imposing the symmetry $\theta \to -\theta$ on the 
perturbations goes 
hand in hand with imposing $\theta \to -\theta$ for the source which means that one should add a corresponding
delta-function $\delta(\theta+\theta')$ to the right hand side of (\ref{TransvGreen}). The 
Israel-condition (\ref{IsraelijFluct}) on the brane with $\theta=0$ then gives 
$\frac{\partial g^{(\lambda)}}{\partial \theta}=0$, as in the Randall-Sundrum case. 

Since in our setup we do not have any convincing argument for imposing the $\theta\to-\theta$ symmetry also on the
metric fluctuations (the scalar field), the second case, which allows a breaking of this symmetry will be the
case of our choice. The Israel-condition (\ref{IsraelijFluct}) now merely indicates that the perturbation
(the scalar field) should have a continuous first derivative at the brane.

In order to obtain a uniquely defined solution we have to impose two boundary conditions 
on $g^{(\lambda)}(\theta,\theta')$ one at $\theta=\pi/2$ the other at $\theta=-\pi/2$. It turns out that
the requirement of square integrability of our solutions (with the correct weight-function) forces us to 
discard one of the two fundamental solutions at the boundaries $\theta=\pm\pi/2$, namely 
$\varphi_2(\lambda,\theta)$.

Before going into details we introduce the following abbreviations:
\begin{align} 
  p(\theta)=\frac{\sigma(\theta)\gamma(\theta)^3}{R^2}\, ,  \qquad 
  s(\theta)=\frac{\sigma(\theta)\gamma(\theta)}{R_U^2} .
\end{align}
Eq. (\ref{TransvGreen}) then becomes
\begin{equation} 
  \left[p(\theta) {g^{(\lambda)}}{'}(\theta,\theta')\right]'-(\lambda^2-1) \,
  s(\theta) g^{(\lambda)}(\theta,\theta')=\delta(\theta-\theta') \, ,
\end{equation}
where $'$ denotes derivatives with respect to $\theta$ here and in the following. 
We start by taking $\theta'>0$. The case $\theta'<0$ is fully analogous to the one considered 
up to minus signs.
Our ansatz for $g^{(\lambda)}(\theta,\theta')$ is
\begin{equation} \label{GreenAnsatz}
  g^{(\lambda)}(\theta,\theta')=\left\{
  \begin{array}{lrrl} A(\theta') \, \varphi_1^{(\lambda)}(-\theta)&-\frac{\pi}{2}&\leq \theta \leq& 0 \, ,\\
                    B(\theta') \, \varphi_1^{(\lambda)}(\theta)+C(\theta') \, \varphi_2^{(\lambda)}(\theta)\quad
                             &0 &\leq \theta \leq& \theta' \, ,\\
                    D(\theta') \, \varphi_1^{(\lambda)}(\theta)&\theta' &\leq \theta \leq& \frac{\pi}{2} \, .
  \end{array} \right.
\end{equation}
We impose the following boundary and matching conditions on $g^{(\lambda)}(\theta,\theta')$ which will uniquely 
determine the coefficients $A, B, C, D$:
\begin{enumerate} 
  \item{continuity of $g^{(\lambda)}(\theta,\theta')$ at $\theta=0$.}
  \item{continuity of ${g^{(\lambda)}}{'}(\theta,\theta')$ at $\theta=0$.}
  \item{continuity of $g^{(\lambda)}(\theta,\theta')$ at $\theta=\theta'$.}
  \item{jump condition of ${g^{(\lambda)}}{'}(\theta,\theta')$ at $\theta=\theta'$.}
\end{enumerate}
Using the ansatz (\ref{GreenAnsatz}) we obtain from the above conditions:
\begin{equation} 
  \begin{array}{rclcc}
  \left[A(\theta')-B(\theta')\right] \varphi_1^{(\lambda)}(0) &-&C(\theta') \varphi_2^{(\lambda)}(0)&=&0\, ,\\
  \left[A(\theta')+B(\theta')\right] {\varphi_1^{(\lambda)}}{'}(0) &+&C(\theta') 
   {\varphi_2^{(\lambda)}}{'}(0)&=&0\, , \\
  \left[B(\theta')-D(\theta')\right] \varphi_1^{(\lambda)}(\theta') &+&C(\theta') 
   \varphi_2^{(\lambda)}(\theta')&=&0\, ,\\
  \left[B(\theta')-D(\theta')\right] {\varphi_1^{(\lambda)}}{'}(\theta') &+&C(\theta') 
  {\varphi_2^{(\lambda)}}{'}(\theta')&=&-\frac{1}{p(\theta')}   \, ,
  \end{array}
\end{equation}
the solution of which is easily found to be:
\begin{align} \label{solABCD}
  A(\theta')&=\frac{R^2}{2}\frac{\varphi_1^{(\lambda)}(\theta')}
   {\varphi_1^{(\lambda)}(0) {\varphi_1^{(\lambda)}}{'}(0)}
  \, ,\quad
  B(\theta')=\frac{R^2}{2}\frac{\varphi_1^{(\lambda)}(\theta')}
         {\mathcal{W}\left[\varphi_1^{(\lambda)},\varphi_2^{(\lambda)},0\right]}
         \frac{\varphi_1^{(\lambda)}(0) {\varphi_2^{(\lambda)}}{'}(0)+
         \varphi_2^{(\lambda)}(0) {\varphi_1^{(\lambda)}}{'}(0)}
         {\varphi_1^{(\lambda)}(0) {\varphi_1^{(\lambda)}}{'}(0)} \, ,\nonumber \\
  C(\theta')&=-R^2 \frac{\varphi_1^{(\lambda)}(\theta')}{\mathcal{W}
          \left[\varphi_1^{(\lambda)},\varphi_2^{(\lambda)},0\right]} \, , \qquad \qquad
  D(\theta')=B(\theta')- R^2 \frac{\varphi_2^{(\lambda)}(\theta')}{\mathcal{W}
          \left[\varphi_1^{(\lambda)},\varphi_2^{(\lambda)},0\right]} \, ,
\end{align}
where $\mathcal{W}\left[\varphi_1^{(\lambda)},\varphi_2^{(\lambda)},0\right]$ denotes the Wronskian 
of $\varphi_1^{(\lambda)}$ and $\varphi_2^{(\lambda)}$ at $\theta=0$. In obtaining (\ref{solABCD}) we 
used the relation
\begin{equation} 
  \mathcal{W}\left[\varphi_1^{(\lambda)},\varphi_2^{(\lambda)},\theta\right]=
  \frac{\mathcal{W}\left[\varphi_1^{(\lambda)},\varphi_2^{(\lambda)},0\right]}{R^2\, p(\theta)} \, .
\end{equation}
As already mentioned, the case $\theta'<0$ can be treated in complete analogy to the case $\theta'>0$. 
Combining the two results gives the final expression for the solution $g^{(\lambda)}(\theta,\theta')$ of 
(\ref{TransvGreen}) (for  $\lambda=2,3,\ldots$): 
\begin{equation} \label{GreenfullResult}
  g^{(\lambda)}(\theta,\theta')=\left\{
  \begin{array}{lrrl} A(\vert \theta'\vert) \, \varphi_1^{(\lambda)}(\vert \theta\vert)&-\frac{\pi}{2}
                         &\leq \theta \, \mbox{sign}(\theta') \leq& 0\, , \\
                 B(\vert \theta'\vert) \, \varphi_1^{(\lambda)}(\vert \theta\vert)+C \, (\vert \theta'\vert) 
                     \varphi_2^{(\lambda)}(\vert \theta\vert)\quad
                         &0 &\leq \theta \, \mbox{sign}(\theta') \leq& \vert \theta' \vert \, , \\
                    D(\vert \theta'\vert) \, \varphi_1^{(\lambda)}(\vert \theta \vert )
                      & \vert \theta' \vert & \leq \theta \, \mbox{sign}(\theta') \leq& \frac{\pi}{2} \, .
  \end{array} \right.
\end{equation}
It is easily verified that the Green's function (\ref{GreenfullResult}) has the property
\begin{equation} 
   g^{(\lambda)}(\theta,\theta') = g^{(\lambda)}(\theta',\theta)
\end{equation}
as expected for a self-adjoint boundary value problem. We will not need the result 
(\ref{GreenfullResult}) in its full generality. We focus on the case where sources are located at
the brane ($\theta'=0$). In this case (\ref{GreenfullResult}) reduces to:
\begin{equation} \label{glambdaratio}
  g^{(\lambda)}(\theta,0)=\frac{R^2}{2}
  \frac{\varphi_1^{(\lambda)}(\vert \theta\vert)}{{\varphi_1^{(\lambda)}}{'}(0)} \, , 
  \qquad (\lambda=2,3,\ldots) \, .
\end{equation}

For the sake of completeness, we also give the solution $g^{(1)}(\theta,\theta')$ to (\ref{TransverseEq1}) 
\begin{equation} \label{Transverse1Repeat}
  \frac{1}{R^2} \frac{1}{\sigma(\theta)\gamma(\theta)^3} \frac{\partial}{\partial\theta}
    \left[ \sigma(\theta)\gamma(\theta)^3 \frac{\partial g^{(1)}(\theta,\theta')}{\partial \theta}
    \right]=\frac{1}{\sigma(\theta)\gamma(\theta)^3} \delta(\theta-\theta')-
    \tilde{\chi}_1(\theta) \bar{\tilde{\chi}}_1(\theta') \, .
\end{equation}
The general procedure of finding the solution is fully analogous to the case of 
$g^{(\lambda)}(\theta,\theta')$ (for $\lambda=2,3,\ldots$), the only difference being 
that in regions where $\theta \neq \theta'$ (\ref{Transverse1Repeat}) reduces to an
inhomogeneous differential equation. In the interval $0 \leq \theta \leq \pi/2$ the general solution
is given by
\begin{eqnarray} 
  \psi(\theta)=\psi_0(\theta) + A \psi_1(\theta)+B \psi_2(\theta),
\end{eqnarray}
with
\begin{align} 
  \psi_0(\theta)&=-\frac{R^2}{2 \omega \tanh\left( \frac{\omega\pi}{2}\right)} 
  \ln\left[ \cosh\left[\omega \left(\frac{\pi}{2}-\theta\right)\right]\right]\, ,\nonumber \\
  \psi_1(\theta)&=1\, , \qquad \psi_2(\theta)=2 \ln\left[\tanh\left[\omega \left(\frac{\pi}{2}-\theta\right)\right]\right]+
   \coth^2\left[\omega \left(\frac{\pi}{2}-\theta\right)\right]. 
\end{align}
With these definitions, it is easy to verify that $g^{(1)}(\theta,\theta')$ is given by:
\begin{equation} \label{resg1}
  g^{(1)}(\theta,\theta')=\mbox{const.}+\psi_0(\vert \theta\vert)+\psi_0(\vert \theta'\vert)+C 
  \left\{\begin{array}{lrcl} \psi_2(0) & -\frac{\pi}{2} \leq \theta \, \mbox{sign}(\theta')\leq & 0 \\
                             \psi_2(\vert \theta\vert) & 0 \leq \theta \, \mbox{sign}(\theta') \leq &\vert \theta' \vert\\
                             \psi_2(\vert \theta' \vert) & \vert \theta' \vert \leq \theta \, \mbox{sign}(\theta') \leq& 
                              \frac{\pi}{2}\end{array} \right. \, ,
\end{equation}
with 
\begin{equation}
   C=-\frac{R^2}{\mathcal{W}\left[\psi_1,\psi_2,0 \right]} \, .
\end{equation}
By $\mathcal{W}\left[\psi_1,\psi_2,0 \right]$ we again mean the Wronskian of $\psi_1$ and $\psi_2$ 
evaluated at $\theta=0$.

\newpage
\section{Detailed computation of the corrections to Newton's law} \label{appSum}
\setcounter{equation}{0}
In this appendix we give a detailed computation of the sum
\begin{equation} \label{TheSumB}
  S[s,\theta,\omega]=\sum_{\lambda=2}^{\infty}\lambda \sin \left(\lambda s \right)
  \frac{\varphi_1^{(\lambda)}(\theta)}{{\varphi'}_1^{(\lambda)}(0)} \, ,\qquad 0<\theta \leq \frac{\pi}{2}
\end{equation}
with 
\begin{eqnarray} \label{RatioOf2Hyps}
  \frac{\varphi_1^{(\lambda)}(\theta)}{{\varphi'}_1^{(\lambda)}(0)}&=& 
  -\frac{z(\theta)^\frac{\lambda-1}{2} \left[1-z(\theta)\right]^2 
  \phantom{1}_2 F_1\left[\frac{\lambda+3}{2},\frac{\lambda+3}{2};\lambda+1;z(\theta)\right]}
  {\omega(\lambda-1) z(0)^\frac{\lambda-2}{2} \left[1-z(0)\right]^2 
  \phantom{1}_2 F_1\left[ \frac{\lambda+1}{2},\frac{\lambda+3}{2};\lambda+1;z(0) \right]} \, ,
\end{eqnarray}
where we explicitly excluded $\theta=0$ since the Fourier sum (\ref{TheSumB}) does
not converge for this value of~$\theta$.\footnote{Note that this mathematical delicacy about the 
divergence of the Green's function on the brane also applies to the Randall-Sundrum case of the 
Fourier integral representation of the two-point function.
The obvious remedy is to keep $\theta$ strictly greater than
zero during the whole calculation and eventually take the limit $\theta \to 0$.}
We also used the definition $z(\theta)=\tanh^2\left[\omega\left(\frac{\pi}{2}-\theta\right)\right]$.

Note that the sum $S[s,\theta,\omega]$ contains the contribution to the two-point function
coming from zero mode and higher Kaluza-Klein modes. Inspired by the Randall-Sundrum 
case we try to separate the two contributions by functional relations between contiguous 
Gauss Hypergeometric functions. Using (15.2.15) of reference \cite{AbrStegun} for 
$a=b-1=\left(\lambda+1\right)/2,\,c=\lambda+1$ we obtain
\begin{eqnarray} 	
  \left[1-z(\theta)\right] \phantom{1}_2 F_1\left[ \frac{\lambda+3}{2},\frac{\lambda+3}{2};
  \lambda+1;z(\theta) \right]&=&
  \frac{2}{\lambda+1}
  \phantom{1}_2 F_1\left[ \frac{\lambda+1}{2},\frac{\lambda+3}{2};
  \lambda+1;z(\theta) \right] \nonumber \\
  &+&\frac{\lambda-1}{\lambda+1}
  \phantom{1}_2 F_1\left[ \frac{\lambda+1}{2},\frac{\lambda+1}{2};
  \lambda+1;z(\theta) \right] \, .
\end{eqnarray}
In this way we are able to rewrite $S[s,\theta,\omega]$ in the form:
\begin{equation} 
  S[s,\theta,\omega]=  S_1[s,\theta,\omega] + S_2[s,\theta,\omega]\, ,
\end{equation}
where
\begin{eqnarray} \label{defS1}
  S_1[s,\theta,\omega]=-\frac{2}{\omega}\frac{z(0)}{z(\theta)^\frac{1}{2}} 
  \frac{1-z(\theta)}{\left[1-z(0)\right]^2}
  \sum_{\lambda=2}^{\infty}\frac{\lambda \sin\left( \lambda s\right)}{\lambda^2-1} 
  \left[\frac{z(\theta)}{z(0)}\right]^\frac{\lambda}{2}
  \frac{\phantom{1}_2 F_1\left[ \frac{\lambda+1}{2},\frac{\lambda+3}{2};
  \lambda+1;z(\theta) \right]}{\phantom{1}_2 F_1\left[ \frac{\lambda+1}{2},\frac{\lambda+3}{2};
  \lambda+1;z(0) \right]}\, , \\
  S_2[s,\theta,\omega]=-\frac{1}{\omega}\frac{z(0)}{z(\theta)^\frac{1}{2}}
  \frac{1-z(\theta)}{\left[1-z(0)\right]^2}
  \sum_{\lambda=2}^{\infty}\frac{\lambda \sin\left( \lambda s\right)}{\lambda+1} \label{defS2}
  \left[\frac{z(\theta)}{z(0)}\right]^\frac{\lambda}{2}
  \frac{\phantom{1}_2 F_1\left[ \frac{\lambda+1}{2},\frac{\lambda+1}{2};
  \lambda+1;z(\theta) \right]}{\phantom{1}_2 F_1\left[ \frac{\lambda+1}{2},\frac{\lambda+3}{2};
  \lambda+1;z(0) \right]} \, .
\end{eqnarray}
From the asymptotic formula (\ref{GeneralAsymptoticOfHyp}) we infer that now the sum $S_1[s,\theta,\omega]$ 
is convergent even for $\theta=0$ due to the additional power of $\lambda$ in the denominator. We therefore
obtain
\begin{equation} 
  S_1[s,0,\omega]=-\frac{2}{\omega} \frac{z(0)^\frac{1}{2}}{1-z(0)}
  \sum_{\lambda=2}^{\infty}\frac{\lambda \sin\left( \lambda s\right)}{\lambda^2-1} 
  =-\frac{2}{\omega} \frac{z(0)^\frac{1}{2}}{1-z(0)}
  \left(\frac{\pi-s}{2}\cos s -\frac{1}{4}\sin s \right)\, .
\end{equation}
Since we recognized in the last sum the two point function of Einstein's static universe, we attribute
the contribution coming from $S_1[s,\theta,\omega]$ to the zero mode
in the Kaluza-Klein spectrum.
We also should mention that in $S_2[s,\theta,\omega]$ we still need $\theta>0$ for convergence since 
only then the factor $\left[z(\theta)/z(0)\right]^\frac{\lambda-1}{2}$ provides an exponential cutoff 
in $\lambda$ for the sum.

The main challenge in the evaluation of the two-point function is therefore to tame the sum 
$S_2[s,\theta,\omega]$. Our strategy is the following: first we extend it from $\lambda=0$ to
$\infty$ by adding and subtracting the $\lambda=0$ and $\lambda=1$ terms. Next, we replace the sum
over $\lambda$ by the sum of two integrals employing a variant of the \textsc{Euler-Maclaurin} sum 
rule called the
\textsc{Abel-Plana} formula (see. e.g. \cite{Olver}, p. 289-290). Finally, we shall see that the
 resulting (exact) integral representation will allow us to extract the desired asymptotic of the 
two-point function on the brane in the distance regime of interest ($R\ll r \ll R_U$). 

As announced, we start by extending the range of the sum from $0$ to $\infty$. Since the addend with
$\lambda=0$ vanishes, we only need to subtract the $\lambda=1$ term with the result: 
\begin{equation} \label{S2extended}
  S_2[s,\theta,\omega]=-\frac{1}{2 \omega}\frac{z(0)^\frac{1}{2}}{z(\theta)} \frac{1-z(\theta)}{1-z(0)} 
  \ln\left[1-z(\theta)\right] \sin s 
  -\frac{1}{\omega} \frac{z(0)}{z(\theta)^\frac{1}{2}} \frac{1-z(\theta)}{1-z(0)}
  R[s,\theta,\omega]\; ,
\end{equation}
where $R[s,\theta,\omega]$ is defined by 
\begin{equation} \label{defR}
  R[s,\theta,\omega]=\sum_{\lambda=0}^{\infty}\frac{\lambda \sin \left(\lambda s\right)}{\lambda+1}
  \left[\frac{z(\theta)}{z(0)}\right]^\frac{\lambda}{2}
  \frac{\phantom{1}_2 F_1\left[ \frac{\lambda+1}{2},\frac{\lambda+1}{2};
  \lambda+1;z(\theta) \right]}{\phantom{1}_2 F_1\left[ \frac{\lambda+1}{2},\frac{\lambda-1}{2};
  \lambda+1;z(0) \right]} \, .
\end{equation}
We used the identity 
\begin{equation} \label{HypIdentity}
  \phantom{1}_2 F_1\left[ \frac{\lambda+1}{2},\frac{\lambda+3}{2};\lambda+1;z(0) \right]=
  \left[1-z(0)\right]^{-1} 
  \phantom{1}_2 F_1\left[ \frac{\lambda+1}{2},\frac{\lambda-1}{2};\lambda+1;z(0) \right]
\end{equation}
in the last step (see 15.3.3 of \cite{AbrStegun}).
The next step consists of extending the $\sin$ function in
(\ref{defR}) to an exponential and taking the imaginary part out of the sum (as in the 
Randall-Sundrum case). Now we make use of the \textsc{Abel-Plana} formula, 
first considering the partial sums:
\begin{eqnarray} 
  R^{(n)}[s,\theta,\omega]&=&\Im \left[ \frac{1}{2} f(0,s,\theta,\omega)+
        \frac{1}{2} f(n,s,\theta,\omega)+\int\limits_0^n f(\lambda,s,\theta,\omega) \, d\lambda \right.  \\
	&&+\imath \left. \int\limits_0^{\infty}\frac{f(\imath y,s,\theta,\omega)-
         f(n+\imath y,s,\theta,\omega)-
        f(-\imath y,s,\theta,\omega)+f(n-\imath y,s,\theta,\omega)}{e^{2\pi y}-1} dy  \right] \, ,\nonumber
\end{eqnarray}
where for the sake of clarity we introduced 
\begin{equation} \label{fdef}
  f(\lambda,s,\theta,\omega)\equiv \frac{\lambda e^{\imath\lambda s}}{\lambda+1}
  \left[\frac{z(\theta)}{z(0)}\right]^\frac{\lambda}{2}
  \frac{\phantom{1}_2 F_1\left[ \frac{\lambda+1}{2},\frac{\lambda+1}{2};
  \lambda+1;z(\theta) \right]}{\phantom{1}_2 F_1\left[ \frac{\lambda+1}{2},\frac{\lambda-1}{2};
  \lambda+1;z(0) \right]}.
\end{equation}
In the limit $n\to\infty$ we find:
\begin{eqnarray} \label{TIntRep}
  R[s,\theta,\omega]&=&\Im \Big[ \underbrace{\int\limits_0^\infty f(\lambda,s,\theta,\omega) d\lambda}_
	{\equiv R_1[s,\theta,\omega]}
	+\underbrace{\imath \int\limits_0^{\infty}\frac{f(\imath y,s,\theta,\omega)-
	f(-\imath y,s,\theta,\omega)}{e^{2\pi y}-1} dy}_{\equiv R_2[s,\theta,\omega]} \Big] \, .
\end{eqnarray}
\subsection{Leading order asymptotic of $\Im\left\{R_1[s,\theta,\omega]\right\}$}
We focus first on $R_1[s,\theta,\omega]$ and observe that the function 
$f(\lambda,s,\theta,\omega)$ is holomorphic (in $\lambda$) in the first 
quadrant.\footnote{Everything but the ratio of the hypergeometric functions is clearly holomorphic. 
The dependence
of Gauss's hypergeometric function on the parameters is also holomorphic so the only danger comes from
the zeros of $\phantom{1}_2 F_1\left[ \frac{\lambda+1}{2},\frac{\lambda-1}{2};\lambda+1;z(0) \right]$. These,
however, turn out to be outside of the first quadrant.}
Therefore, in perfect analogy with the Randall-Sundrum case treated in appendix~\ref{appParallel}, 
we can use \textsc{Cauchy}'s theorem to replace the integration over the positive real 
axis by an 
integration over the positive imaginary axis.\footnote{To be mathematically correct, one has to apply
\textsc{Cauchy}'s theorem to the quarter of the disk of radius $R$  bounded by the positive real axis, the 
positive imaginary axis and the arc joining the points $R$ and $\imath R$, see Fig.~\ref{intcontour} in 
appendix \ref{appParallel}.
The above statement will then be correct if in the limit $R\to \infty$ the contribution from 
the arc tends to zero. We will verify this in section \ref{arcjust} of this appendix.} 
After substituting $\lambda$ with $\imath y$ we obtain :
\begin{equation} \label{R1shifted}
  R_1[s,\theta,\omega]=-\int\limits_0^\infty \frac{y (1-\imath y)}{1+y^2} e^{-y s}
  \left[\frac{z(\theta)}{z(0)}\right]^\frac{\imath y}{2}
  \frac{\phantom{1}_2 F_1\left[ \frac{1+\imath y}{2},\frac{1+\imath y}{2};
  1+\imath y;z(\theta) \right]}{\phantom{1}_2 F_1\left[ \frac{1+\imath y}{2},\frac{-1+\imath y}{2};
  1+\imath y;z(0) \right]} dy \, .
\end{equation}
The upper deformation of the path of integration is advantageous for at least two reasons. Firstly, 
through the appearance of the exponential in the integrand, convergence on the brane does no longer rely
on $\theta>0$ and we can set $\theta$ equal to zero in (\ref{R1shifted}) to obtain:
\begin{equation} \label{R1shiftedBrane}
  R_1[s,0,\omega]=-\int\limits_0^\infty \frac{y (1-\imath y)}{1+y^2} e^{-y s}
  \frac{\phantom{1}_2 F_1\left[ \frac{1+\imath y}{2},\frac{1+\imath y}{2};
  1+\imath y;z(0) \right]}{\phantom{1}_2 F_1\left[ \frac{1+\imath y}{2},\frac{-1+\imath y}{2};
  1+\imath y;z(0) \right]} dy \, .
\end{equation}   
Secondly, the form (\ref{R1shifted}) is perfectly suited for obtaining asymptotic expansions
for large distances. Note that since $s\in[0,\pi]$, we carefully avoided saying 
``for large s'' since this would not correspond to the case of our interest. We try to compute
the two-point function in a distance regime $R \ll r \ll R_U$, distances clearly far beyond 
the size of the fifth dimension but far below the size of the observable universe. Rewritten in
the geodesic distance coordinate $s$ this becomes 
$\omega/\sinh\left(\frac{\omega\pi}{2}\right)\ll s\ll 1$. It is this last relation which complicates
the computation of the asymptotic considerably. We are forced to look at an asymptotic 
evaluation of (\ref{R1shiftedBrane}) (or (\ref{R1shifted})) for intermediate $s$ values such that
 we cannot make direct use of Laplace's method. The correct way of extracting the above described
asymptotic is to expand the ratio of hypergeometric functions in (\ref{R1shiftedBrane}) in a power 
series of the small parameter $1-z(0)$. Using rel. $(15.3.10)$ and $(15.3.11)$ (with $m=1$) of 
\cite{AbrStegun} we find after some algebra
\begin{eqnarray} \label{HypRatioExpansion}
  &&\frac{\phantom{1}_2 F_1\left[ \frac{1+\imath y}{2},\frac{1+\imath y}{2};
  1+\imath y;z(0) \right]}{\phantom{1}_2 F_1\left[ \frac{1+\imath y}{2},\frac{-1+\imath y}{2};
  1+\imath y;z(0) \right]}= 
  -\frac{1}{2}(1+\imath y)\left[2\gamma+2\psi(\frac{1+\imath y}{2})+\ln\left[1-z(0)\right]\right]  
	\hfill \nonumber \\
  &+&\left\{\left(\frac{1+\imath y}{2}\right)^3\left[2 \psi(2)-2 \psi\left(\frac{3+\imath y}{2}\right)
     -\ln\left[1-z(0)\right]\right] \right.\nonumber \\
  &&\;\;\;-\left(\frac{1+\imath y}{2}\right)^2 \left(\frac{-1+\imath y}{2}\right)
    \left[2 \psi(1)-2 \psi\left(\frac{1+\imath y}{2}\right)-\ln\left[1-z(0)\right]\right]\times\nonumber \\
  &&\;\;\;\;\;\;\;\left.\times \left[\ln\left[1-z(0)\right]-\psi(1)-\psi(2)
     +\psi\left(\frac{3+\imath y}{2}\right)+
     \psi\left(\frac{1+\imath y}{2}\right)\right] \right\} \left[1-z(0)\right]\nonumber\\
  &+&\mathcal{O}\left\{\ln\left[1-z(0)\right] \left[1-z(0)\right]^2 \right\} \, .
\end{eqnarray}
In the above relation $\gamma$ denotes \textsc{Euler-Mascheroni}'s constant and  
$\psi(z)\equiv\Gamma'[z]/\Gamma[z]$ the so-called Digamma-function. Note that we 
developed up to linear order in
$1-z(0)$ since we also want to compute the next to leading order term in the asymptotic later
in this appendix. 

We are only interested in the imaginary part of $R_1[s,0,\omega]$ and it turns out that the first term in
the expansion (\ref{HypRatioExpansion}) after insertion in (\ref{R1shiftedBrane}) can be integrated 
analytically:
\begin{eqnarray} \label{R1_0_Integral}
  \Im\left\{R_1^{(0)}[s,0,\omega]\right\}=\Im \left\{\int\limits_0^{\infty} y e^{-y s} 
  \psi\left(\frac{1+\imath y}{2}\right)
  dy \right\}=\int\limits_0^{\infty} y e^{-y s} \Im\left[\psi\left(\frac{1+\imath y}{2}\right)\right] dy \, .
\end{eqnarray}
The last step is justified since also the real part of the first integral in (\ref{R1_0_Integral}) 
converges, as follows 
immediately from the asymptotic $\psi(z)\sim\ln z-\frac{1}{2 z}+\mathcal{O}(\frac{1}{z^2})$. Since
\begin{eqnarray} \label{ImofDiGamma}
  \Im\left[\psi(\frac{1+\imath y}{2})\right]=\frac{\pi}{2}\tanh \left(\frac{\pi y}{2}\right) \, ,
\end{eqnarray}
(\cite{AbrStegun}, p.259, 6.3.12) we find
\begin{eqnarray} \label{R1_0Result}
  \Im\left\{R_1^{(0)}[s,0,\omega]\right\}&=&\frac{\pi}{2}\int\limits_{0}^\infty y e^{-y s} 
   \tanh \left(\frac{\pi y}{2}\right) dy
   =-\frac{\pi}{2} \frac{d}{ds}\left[\int\limits_{0}^\infty e^{-y s} 
   \tanh\left(\frac{\pi y}{2}\right) dy\right] \nonumber \\
   &=&-\frac{\pi}{2} \frac{d}{ds}\left[\int\limits_{0}^\infty e^{-y s} \left( 
    \frac{2}{1+e^{-\pi y}}-1\right) dy\right] =
    -\frac{\pi}{2}\frac{d}{ds}\left[-\frac{1}{s}+2\int\limits_0^\infty \frac{e^{-y s}}{1+e^{-\pi y}} dy \right] 
    \nonumber \\
   &=&-\frac{\pi}{2 s^2}-\frac{d}{ds}\left[\int\limits_0^\infty
   \frac{e^{-\frac{z s}{\pi}}}{1+e^{-z}} dz\right]=-\frac{\pi}{2 s^2}-\frac{1}{\pi}\beta'(\frac{s}{\pi})\, , 
\end{eqnarray}
where we introduced the $\beta$-function by
\begin{equation} \label{defbeta}
  \beta(x)\equiv\frac{1}{2}\left[\psi\left(\frac{x+1}{2}\right)-\psi\left(\frac{x}{2}\right)\right]
\end{equation}
(see \cite{Gradshteyn}, p.331, 3.311, 2.).\footnote{By $\beta'$ in (\ref{R1_0Result}) we mean the derivative of $\beta$ with respect to its argument.}

\subsection{Leading order asymptotic of $\Im\left\{R_2[s,\theta,\omega]\right\}$}
We can treat $R_2[s,\theta,\omega]$ starting from (\ref{TIntRep}) in very much the same way as 
$R_1[s,\theta,\omega]$. Noting that we can again put $\theta$ equal to zero since the exponential 
factor $e^{2\pi y}$ guarantees the convergence of the integral, $R_2[s,0,\omega]$ becomes explicitly:
\begin{eqnarray} \label{R2explicit}
  R_2[s,0,\omega]=\int\limits_0^\infty \frac{1}{e^{2\pi y}-1}&&\left[ \frac{-y+\imath y^2}{1+y^2}
   e^{-y s} \frac{\phantom{1}_2 F_1\left[ \frac{1+\imath y}{2},\frac{1+\imath y}{2};
  1+\imath y;z(0) \right]}{\phantom{1}_2 F_1\left[ \frac{1+\imath y}{2},\frac{-1+\imath y}{2};
  1+\imath y;z(0) \right]} \right. \nonumber \\
   && \left. +\frac{-y-\imath y^2}{1+y^2}
   e^{y s} \frac{\phantom{1}_2 F_1\left[ \frac{1-\imath y}{2},\frac{1-\imath y}{2};
  1-\imath y;z(0) \right]}{\phantom{1}_2 F_1\left[ \frac{1-\imath y}{2},\frac{-1-\imath y}{2};
  1-\imath y;z(0) \right]} \right] dy \, .
\end{eqnarray}
If we now employ the expansion (\ref{HypRatioExpansion}) two times in (\ref{R2explicit}) we find 
after some algebra:
\begin{eqnarray}
  \Im\left\{R_2^{(0)}[s,0,\omega]\right\}=\Im \left\{ \int\limits_0^\infty \frac{y}{e^{2\pi y}-1} \left[
  e^{-y s} \psi\left(\frac{1+\imath y}{2}\right)+e^{y s} \psi\left(\frac{1-\imath y}{2}\right) \right] 
  dy \right\}
\end{eqnarray}
and using (\ref{ImofDiGamma}) we have
\begin{align} \label{R2IntRes}
  \Im\left\{R_2^{(0)}[s,0,\omega]\right\}&=\int\limits_0^\infty \frac{\pi}{2} \frac{y}{e^{2\pi y}-1}
  \left[ e^{-y s} \tanh \left(\frac{\pi y}{2}\right)+e^{y s} \tanh \left(-\frac{\pi y}{2}\right) \right] dy\\
  &=-\pi\int\limits_0^\infty \frac{y}{e^{2\pi y}-1}\tanh\left(\frac{\pi y}{2}\right)\sinh(y s)dy=
  -\pi\int\limits_0^\infty \frac{y}{\left(1+e^{\pi y}\right)^2}\sinh (y s) dy\nonumber \\
  &=-\pi\frac{d}{ds}\left[\int\limits_0^\infty \frac{\cosh(y s)}{\left(1+e^{\pi y}\right)^2} dy\right]=
  -\frac{1}{2}\frac{d}{ds}\left[\int\limits_0^1\frac{u^{1-\frac{s}{\pi}}}{(1+u)^2} du +
                                \int\limits_0^1\frac{u^{1+\frac{s}{\pi}}}{(1+u)^2} du \right] \nonumber \\
  &=-\frac{1}{2}\frac{d}{ds} \left\{\frac{1}{2-\frac{s}{\pi}} 
        \phantom{1}_2 F_1\left[2,2-\frac{s}{\pi};3-\frac{s}{\pi};-1 \right]+
	\frac{1}{2+\frac{s}{\pi}} 
        \phantom{1}_2 F_1\left[2,2+\frac{s}{\pi};3+\frac{s}{\pi};-1 \right]\right\} \, .\nonumber 
\end{align}
While we substituted $y$ by $u$ according to $u=e^{-\pi y}$ in the third line, we used (3.194, p.313) of
\cite{Gradshteyn} with $\nu=2$, $u=\beta=1$ and $\mu=2\mp s/\pi$ (valid for $s<2 \pi$)
in the last line.
In order to simplify the hypergeometric functions, we first use another relation between 
contiguous functions, namely eq. (15.2.17) of \cite{AbrStegun} with $a=1$, $b=k$, $c=k+1$ and $z=-1$ and 
then the formulae (15.1.21) and (15.1.23) of \cite{AbrStegun} together with the duplication formula 
for the $\Gamma$-function to obtain:
\begin{eqnarray}
  \frac{1}{k}\phantom{1}_2 F_1\left[2,k;k+1;-1 \right]&=&\phantom{1}_2 F_1\left[1,k;k;-1 \right]-
  \frac{k-1}{k} \phantom{1}_2 F_1\left[1,k;k+1;-1 \right] \nonumber \\
  &=&2^{-k} \sqrt{\pi}\frac{\Gamma\left[k\right]}{\Gamma\left[\frac{k}{2}\right]
  \Gamma\left[\frac{k+1}{2}\right]}-\left(k-1\right) \beta(k)\nonumber \\
  &=&\frac{1}{2}+(1-k)\beta(k).
\end{eqnarray}
After making use of this in (\ref{R2IntRes}) we finally obtain for the contribution to lowest 
order in $1-z(0)$:
\begin{eqnarray} 
  \Im\left\{R_2^{(0)}[s,0,\omega]\right\}=\frac{1}{2}\frac{d}{ds}\left[\left(1-\frac{s}{\pi}\right)
  \beta(2-\frac{s}{\pi})+\left(1+\frac{s}{\pi}\right) \beta(2+\frac{s}{\pi}) \right].
\end{eqnarray}
Summarizing the results to lowest order in $1-z(0)$ we therefore have
\begin{eqnarray} \label{0orderResult}
  \Im\left\{R_1^{(0)}[s,0,\omega]\right\}&=&-\frac{\pi}{2 s^2}-\frac{1}{\pi}\beta'(\frac{s}{\pi}),\nonumber\\
  \Im\left\{R_2^{(0)}[s,0,\omega]\right\}&=&\frac{1}{2}\frac{d}{ds}\left[\left(1-\frac{s}{\pi}\right)
  \beta(2-\frac{s}{\pi})+\left(1+\frac{s}{\pi}\right) \beta(2+\frac{s}{\pi}) \right].
\end{eqnarray}
Expanding (\ref{0orderResult}) around $s=0$ we find:
\begin{align}
  \Im\left\{R_1^{(0)}[s,0,\omega]\right\}&=\frac{\pi}{2 s^2}-\frac{\pi}{12}+\mathcal{O}(s) \, , \nonumber \\
  \Im\left\{R_2^{(0)}[s,0,\omega]\right\}&=\left( \frac{1}{6}-\frac{3\zeta(3)}{2\pi^2}\right) s +
  \mathcal{O}(s^3) \, .
\end{align}
with $\zeta$ denoting \textsc{Riemann}'s $\zeta$-function.
As expected the contribution from $R_2^{(0)}[s,0,\omega]$ is sub-leading with respect to the one 
from $R_1^{(0)}[s,0,\omega]$.

\subsection{Next to leading order asymptotic of $\Im\left\{R_1[s,\theta,\omega]\right\}$}
We would now like to obtain the next to leading order term in the asymptotic of 
$\Im\left\{R_1[s,\theta,\omega]\right\}$, that is the term obtained from (\ref{R1shiftedBrane})
by taking into account the linear contribution in $1-z(0)$ of the expansion (\ref{HypRatioExpansion}).
Since the calculation of the integral obtained in this way is rather cumbersome and anyway 
we do not need the full $s$-dependence, we do not intend to evaluate it fully. Extracting the leading 
$s$-divergence at $s=0$ is sufficient for our purposes here. To start, we insert the linear
 term of 
(\ref{HypRatioExpansion}) in (\ref{R1shiftedBrane}):
\begin{align}
  \Im\left\{R_1^{(1)}[s,0,\omega]\right\}&=
  -\Im\left\{\int\limits_0^\infty dy y \frac{1-\imath y}{1+y^2} e^{-y s}
   \left\{ \left(\frac{1+\imath y}{2}\right)^3 \left[2 \psi(2)-2 \psi\left(\frac{3+\imath y}{2}
   \right) -\ln\left[1-z(0)\right]\right]\right. \right. \nonumber \\
  &- \left(\frac{1+\imath y}{2}\right)^2 \left(\frac{-1+\imath y}{2}\right)
  \left[2 \psi(1)-2 \psi\left(\frac{1+\imath y}{2}\right)-\ln\left[1-z(0)\right]\right]\times \\
  &\times\left.\left. \left[\ln\left[1-z(0)\right]-\psi(1)-\psi(2)+\psi\left(\frac{3+\imath y}{2}\right)+
     \psi\left(\frac{1+\imath y}{2}\right)\right] \right\} \left[1-z(0)\right] \right\} \, .\nonumber
\end{align}
After some algebra and by using the functional relation of the Digamma-function $\psi(z)$ 
(see \cite{AbrStegun})
\begin{equation} 
  \psi\left(z+1\right)=\psi\left(z\right)+\frac{1}{z}
\end{equation}
we find
\begin{eqnarray} \label{R1_1storder}
  \Im\left\{R_1^{(1)}[s,0,\omega]\right\}=-\frac{1-z(0)}{2} \int\limits_0^{\infty}y e^{-y s} \Im \left\{
  \ldots\right\} dy
\end{eqnarray}
with
\begin{eqnarray} \label{R1_1storderIntegrand}
  \Im \left\{ \ldots\right\} &=&\Im\left\{\left(\frac{1+\imath y}{2}\right)^2 
  \left[2 \psi(2)-2 \psi\left(\frac{1+\imath y}{2}\right)-\ln\left[1-z(0)\right]\right]-(1+\imath y) \right.
  \nonumber \\
  &&\qquad+\frac{1+y^2}{4}\left[2\psi\left(1\right)-2\psi\left(\frac{1+\imath y}{2}\right)-
    \ln\left[1-z(0)\right]\right]\times\nonumber\\
  &&\qquad\quad\qquad\times\left[\ln\left[1-z(0)\right]-\psi\left(1\right)-\psi\left(2\right)+
    2\psi\left(\frac{1+\imath y}{2}\right)\right] \nonumber \\
  &&\qquad+\left. \frac{1-\imath y}{2}\left[2\psi\left(1\right)-2\psi\left(\frac{1+\imath y}{2}\right)
    -\ln\left[1-z(0)\right]\right] \right\} \, .
\end{eqnarray}
We will now keep only the terms proportional to $y^3$ in (\ref{R1_1storder}) 
($y^2$ in (\ref{R1_1storderIntegrand})) since only these will contribute to the leading
$1/s^{4}$ singular behavior. After a few lines of algebra we find
\begin{eqnarray} \label{R1_1HighestContribution}
  \Im\left\{R_1^{(1)}[s,0,\omega]\right\}&=&-\frac{1-z(0)}{2}\left[1-2\gamma-\ln\left[1-z(0)\right]\right]
    \int\limits_0^{\infty} y^3 e^{-y s} \Im \left[\psi\left(\frac{1+\imath y}{2}\right)\right] dy \nonumber\\ 
  &&+\left[1-z(0)\right] \int\limits_0^\infty y^3 e^{-y s} 
    \Im\left[\psi\left(\frac{1+\imath y}{2}\right)\right] 
    \Re\left[\psi\left(\frac{1+\imath y}{2}\right)\right] dy \nonumber \\
  && + \mbox{terms involving lower powers of }y.
\end{eqnarray}
The first integral can be reduced to the integral (\ref{R1_0_Integral}) simply by replacing each power 
of $y$ by a derivative with respect to $s$ and by taking the derivatives out of the integral: 
\begin{eqnarray} 
  \int\limits_0^{\infty} y^3 e^{-y s} \Im \left[\psi\left(\frac{1+\imath y}{2}\right)\right] dy&=&
  -\frac{d^3}{ds^3}\left\{\int\limits_0^\infty e^{-y s} \Im \left[\psi\left(\frac{1+\imath y}{2}\right)\right]
  dy \right\}\nonumber \\
  &=&-\frac{\pi}{2}\frac{d^3}{ds^3}\left[-\frac{1}{s}+\frac{2}{\pi}\beta\left(\frac{s}{\pi}\right)\right] .
\end{eqnarray}
We could not find an analytic expression for the second integral. However all we need is the 
first term of its small $s$  asymptotic\footnote {In the sense $s\ll 1$.} and this can be easily obtained
 from the large $y$ asymptotic of $\Re\left[\psi\left(\left(1+\imath y\right)/2\right)\right]$. Since
asymptotically 
\begin{equation} 
  \psi\left(z\right)\sim\ln z -\frac{1}{2 z}-\frac{1}{12 z^2}+\mathcal{O}(\frac{1}{z^4})\, , \qquad 
  (z \to \infty \; \mbox{in} \; \vert \arg z\vert<\pi)
\end{equation}
we expect that this term will contribute logarithmic terms in $s$ and we find after straightforward
expansions
\begin{eqnarray} \label{RePartAssympt}
  \Re\left[\psi\left(\frac{1+\imath y}{2}\right)\right] \sim \ln\frac{y}{2}-\frac{1}{6 y^2}+
  \mathcal{O}\left(\frac{1}{y^4}\right) \, .
\end{eqnarray}
From the last formula (\ref{RePartAssympt}) we learn that within our current approximation of keeping 
only highest (cubic) power terms in $y$, it is sufficient to take into account only the 
contribution coming from $\ln\frac{y}{2}$. Therefore, we need to calculate the following integral
\begin{equation}
  \frac{\pi}{2} \int\limits_0^\infty y^3 e^{-y s} \tanh \left(\frac{\pi y}{2} \right)\ln y \, dy=
  -\frac{\pi}{2}\frac{d^3}{ds^3}\Big[\underbrace{\int\limits_0^\infty e^{-y s} 
   \tanh \left(\frac{\pi y}{2} \right)
  \ln y \, dy}_{\equiv \mathcal{J}(s)} \Big] \, ,
\end{equation}
where we used (\ref{ImofDiGamma}) once more. 
Note that the term proportional to $\ln 2$ can be accounted for by adding a contribution
of the type of the first integral in (\ref{R1_1HighestContribution}).
One way to find the asymptotic of $\mathcal{J}(s)$ for small $s$ is to integrate the following
asymptotic expansion of $\tanh x$ 
\begin{eqnarray} \label{tanhexp}
  \tanh x = 1-2 e^{-2 x}+2 e^{-4 x}-2 e^{-6 x}+\ldots=1+2 \, \sum_{\nu=1}^\infty (-1)^\nu e^{-2 \nu x}  
\end{eqnarray}
which converges for all $x>0$. Inserting this in the definition of $\mathcal{J}(s)$ we find:
\begin{eqnarray} \label{calIIntRes}
  \mathcal{J}(s)&=&\int\limits_0^\infty e^{-y s} \ln y \, 
  \left[1+2\sum_{\nu=1}^\infty (-1)^\nu e^{-\pi y \nu} \right] dy \qquad \qquad \qquad \qquad
  \qquad \qquad \qquad \qquad \nonumber \\
  &=&-\frac{\gamma + \ln s}{s} +2 \sum_{\nu=1}^\infty (-1)^{\nu+1} 
      \frac{\gamma+\ln\left(s+\nu \pi\right)}{s+\nu\pi}\nonumber\\
  &=& -\frac{\gamma + \ln s}{s} +\frac{2}{\pi}\left(\gamma+\ln\pi\right) \beta\left(\frac{s}{\pi}+1\right)
      +\frac{2}{\pi} \sum_{\nu=1}^\infty (-1)^{\nu+1}
    \frac{\ln\left(\nu+\frac{s}{\pi}\right)}{\nu+\frac{s}{\pi}},
\end{eqnarray}
where we used the series expansion of the $\beta$-function given e.g. in \cite{Gradshteyn}.
The last two terms in the above formula are finite in the limit $s\to0$ 
and therefore will play no role in the asymptotic (since they are multiplied by a factor of $1-z(0)$).

We are now ready to assemble all contributions to the small $s$ asymptotic of 
$\Im\left\{R_1^{(1)}[s,0,\omega]\right\}$:
\begin{align} \label{1OrderResult}
  &\Im\left\{R_1^{(1)}[s,0,\omega]\right\}\sim\frac{1-z(0)}{2}
   \left[1-2\gamma-\ln\left[1-z(0)\right]+2\ln 2\right]
   \frac{\pi}{2}\frac{d^3}{ds^3}\left[-\frac{1}{s}+\frac{2}{\pi}\beta\left(\frac{s}{\pi}\right)\right]
   \nonumber \\
  &\quad -\left[ 1-z(0) \right] \frac{\pi}{2}\frac{d^3}{ds^3}\left[-\frac{\gamma+\ln s}{s}
    + \frac{2}{\pi}\left(\gamma+\ln\pi\right) \beta\left(\frac{s}{\pi}+1\right)
      +\frac{2}{\pi} \sum_{\nu=1}^\infty
    \frac{ (-1)^{\nu+1}\ln\left(\nu+\frac{s}{\pi}\right)}{\nu+\frac{s}{\pi}} \right] 
    +\mathcal{O}(\frac{\ln s}{s^3}) \nonumber \\
  &\qquad= \frac{\pi}{2}\frac{1-z(0)}{s^4} \left\{ 8-6\ln 2 -6 
  \ln\left[\frac{s}{\sqrt{1-z(0)}}\right]\right\} +\mathcal{O}(\frac{\ln s}{s^3}),
\end{align}
where we used the following series expansions for the $\beta$-function and for 
the last term in (\ref{calIIntRes}):
\begin{align}
  \beta\left(\frac{s}{\pi}\right)&=\frac{\pi}{s}-\ln 2 +\mathcal{O}(s)\, ,\nonumber \\
  \beta\left(\frac{s}{\pi}+1\right)&=\ln 2+\mathcal{O}(s)\, ,\nonumber \\
  \sum_{\nu=1}^\infty
    \frac{(-1)^{\nu+1}\ln\left(\nu+\frac{s}{\pi}\right)}{\nu+\frac{s}{\pi}} 
    &=\frac{1}{2} \left[\ln(2)\right]^2 - \gamma \ln(2) + \mathcal{O}(s).
\end{align}
Note that only the term $\beta\left(s/\pi\right)$ gave a contribution to the leading 
$1/s^4$ divergence.
Relations (\ref{0orderResult}) and (\ref{1OrderResult}) constitute the main result of this 
appendix. They provide the desired asymptotic behavior of the two-point function in the regime of 
intermediate distances $\omega/\sinh\left(\frac{\omega\pi}{2}\right) \ll s\ll 1$.
A couple of remarks are in place. Clearly, (\ref{0orderResult}) is not only an asymptotic result but 
is valid for all $s \in [0,\pi]$. Secondly, we were writing $\mathcal{O}(\ln s/s^3)$ in 
(\ref{1OrderResult}) to indicate that we dropped all contributions coming from terms in the integrand 
(\ref{R1_1storderIntegrand}) lower than cubic order. Finally we point out that the development of 
the ratio (\ref{HypRatioExpansion}) in power of $1-z(0)$ indeed generates an asymptotic in the correct 
distance regime. This is clear from the fact that each new power in $1-z(0)$ is paired with an 
additional power of $y^2$ in the integrand of $\Im\left\{R_1[s,\theta,\omega]\right\}$. This by itself
implies upon multiplication by $e^{-y s}$ and integration over $y$ an additional power of 
$1/s^2$. We therefore effectively generate an expansion in powers of 
$\sqrt{1-z(0)}/s=1/\left[s \cosh \left(\frac{\omega\pi}{2}\right)\right]$. It
is also easy to understand why this rough way of counting powers works. The reason is that the 
asymptotic properties of our integrals $\Im\left\{R_1^{(n)}[s,0,\omega]\right\}$ are determined 
only by the behavior of the corresponding integrands for large values of the integration variable $y$.
The terms $\Re\left[\psi\left(\left(1+\imath y\right)/2\right)\right]$ and 
$\Im\left[\psi\left(\left(1+\imath y\right)/2\right)\right]$ do not interfere with the above power counting
since their behavior for large $y$ is either logarithmic in $y$ (for $\Re$) or constant in $y$ (for $\Im$). 

Finally we want to point out that the expansion in powers of 
$1/\left[s \cosh \left(\frac{\omega\pi}{2}\right)\right]$ is
not useful for obtaining information about the Green's function for distances 
$s \ll \omega/\cosh\left(\frac{\omega\pi}{2}\right)$. In this limit the asymptotic expansion of the 
hypergeometric functions in (\ref{RatioOf2Hyps}) for large values of $\lambda$ proves the 
most efficient way to recover the properties of a Green's function in a $4$-dimensional space. 
See appendix \ref{appGreenUltraShort} for the detailed form of the Green's function at distances 
smaller than the extra dimension.

\subsection{Estimate of the arc contribution to $R_1[s,\theta,\omega]$} \label{arcjust}
This final subsection is dedicated to the verification that the arc denoted $C_2$ in 
Fig.~\ref{intcontour} of appendix \ref{appParallel} gives a vanishing contribution to the integral
$R_1[s,\theta,\omega]$ and thus justifying the representation (\ref{R1shiftedBrane}).
We need to know the large $\lambda$ asymptotic behavior of the hypergeometric functions entering 
the definition of $R_1[s,\theta,\omega]$ in (\ref{fdef}). The relevant formula 
(\ref{GeneralAsymptoticOfHyp}) can be found in appendix \ref{appGreenUltraShort} where we 
discuss the short distance behavior of the Green's function. After canceling all common factors from the
ratio of the two hypergeometric functions we find:
\begin{align} \label{FFratioInR1asymptotic}
  \frac{\phantom{1}_2 F_1\left[ \frac{\lambda+1}{2},\frac{\lambda+1}{2};
  \lambda+1;z(\theta) \right]}{\phantom{1}_2 F_1\left[ \frac{\lambda+1}{2},\frac{\lambda-1}{2};
  \lambda+1;z(0) \right]}&=
  \frac{\lambda+1}{\lambda-1} \left[ \frac{z(\theta)}{z(0)}\right]^{-\frac{\lambda+1}{2}} 
  \left[\frac{e^{-\nu(\theta)}}{e^{-\nu(0)}}\right]^\frac{\lambda+1}{2}
  \frac{\left[1+e^{-\nu(\theta)}\right]^{-\frac{1}{2}} \left[1-e^{-\nu(\theta)}\right]^{-\frac{1}{2}}}
  {\left[1+e^{-\nu(0)}\right]^{\frac{3}{2}} \left[1-e^{-\nu(0)}\right]^{\frac{1}{2}}} \times \nonumber \\
  &\qquad \times \left[1+\mathcal{O}\left(\frac{1}{\lambda}\right)\right] \, .
\end{align} 
Substituting $\lambda=R e^{\imath \varphi}$, we now estimate $R_1[s,\theta,\omega]$:
\begin{align} 
  &\left| \int_{C_2} \frac{\lambda e^{\imath\lambda s}}{\lambda+1}
  \left[\frac{z(\theta)}{z(0)}\right]^\frac{\lambda}{2}
  \frac{\phantom{1}_2 F_1\left[ \frac{\lambda+1}{2},\frac{\lambda+1}{2};
  \lambda+1;z(\theta) \right]}{\phantom{1}_2 F_1\left[ \frac{\lambda+1}{2},\frac{\lambda-1}{2};
  \lambda+1;z(0) \right]} d \lambda\right| \nonumber \\
  & \leq \int_{C_2} \left| \frac{\lambda e^{\imath\lambda s}}{\lambda+1}
  \left[\frac{z(\theta)}{z(0)}\right]^\frac{\lambda}{2}
  \frac{\phantom{1}_2 F_1\left[ \frac{\lambda+1}{2},\frac{\lambda+1}{2};
  \lambda+1;z(\theta) \right]}{\phantom{1}_2 F_1\left[ \frac{\lambda+1}{2},\frac{\lambda-1}{2};
  \lambda+1;z(0) \right]} \right| d \lambda \nonumber \\
  & \leq \underbrace{C \left[ \frac{z(0)}{z(\theta)}\right]^{\frac{1}{2}}
  \left| \frac{e^{-\nu(\theta)}}{e^{-\nu(0)}} \right|^\frac{1}{2}
  \frac{\left[1+e^{-\nu(\theta)}\right]^{-\frac{1}{2}} 
  \left[1-e^{-\nu(\theta)}\right]^{-\frac{1}{2}}}{\left[1+e^{-\nu(0)}\right]^{\frac{3}{2}} 
  \left[1-e^{-\nu(0)}\right]^{\frac{1}{2}}}}_{K} R
  \int\limits_0^{\frac{\pi}{2}} 
  \left| e^{\imath R s e^{\imath \varphi}} \right|
  \left| \frac{R e^{\imath \varphi}}{R e^{\imath \varphi -1}} \right| 
  \left| \left[\frac{e^{-\nu(\theta)}}{e^{-\nu(0)}}\right]^{\frac{R e^{\imath \varphi}}{2}}\right| 
   d\varphi \nonumber \\
  &=K R \int\limits_0^{\frac{\pi}{2}} 
  \frac{e^{-R\left(s\sin\varphi-\ln a \cos \varphi\right)}}{\sqrt{1-\frac{2 \cos \varphi}{R}+\frac{1}{R^2}}} 
   d\varphi
  \leq K' R \int\limits_0^{\frac{\pi}{2}} e^{-R s\left(\sin\varphi+\frac{\ln a^{-1}}{s}
   \cos \varphi\right)} d\varphi \, ,
\end{align}
where $C$ is a constant independent of $R$ and the estimate of the ratio of the two hypergeometric 
functions is valid for large enough $R$. 
We also introduced $a$ as in appendix \ref{appGreenUltraShort} by (\ref{adef}). Note that $a<1$ for 
$0<\theta$. In the last line we estimated the inverse of the square root by an arbitrary constant (e.g. 2),
certainly good for large enough $R$ and for all $\varphi$.
All what is left is to observe that
\begin{equation} 
  0 < \epsilon \equiv \min[1,\frac{\ln a^{-1}}{s}] \leq \sin\varphi+\frac{\ln a^{-1}}{s} \cos\varphi \quad
  \mbox{for} \quad s>0, \;\; a^{-1} >1 
\end{equation}
to obtain the desired result:
\begin{equation} 
  \left| \int_{C_2} d\lambda \ldots \right| \leq K' R \int\limits_0^{\frac{\pi}{2}}e^{-R s \epsilon}d\varphi=
  \frac{\pi}{2} K' R e^{-R s \epsilon} \to 0 \quad \mbox{for} \quad R \to \infty.
\end{equation}
 
\newpage
\section{The two point function at very short distances ($r \ll R$)} \label{appGreenUltraShort}
\setcounter{equation}{0}
The aim of this appendix is to calculate the behavior of the two-point function at distances
smaller than the extra dimension $r\ll R$. We will do so by using an appropriate asymptotic expansion 
for large parameter values of the hypergeometric functions in (\ref{RatioOf2Hyps}). Such an asymptotic can be 
found for example in \cite{Erdelyi} or \cite{Luke}. 
The general formula is\footnote{Since we noticed that the formulae given by Luke, p. 452 eq.~(20)--(23) 
\cite{Luke} and the formula given by Erd\'elyi \cite{Erdelyi} differ by a factor of $2^{a+b}$ 
we performed a numerical check. Its outcome clearly favored Erd\'elyi's formula which we reproduced
in \ref{GeneralAsymptoticOfHyp}.}
\begin{align} \label{GeneralAsymptoticOfHyp}
  \phantom{1}_2 F_1\left[a+n,a-c+1+n;a-b+1+2 n;z \right] &= \frac{2^{a+b} \Gamma[a-b+1+2 n]
   \left(\frac{\pi}{n}\right)^{\frac{1}{2}}}{\Gamma[a-c+1+n]\Gamma[c-b+n]}
  \frac{e^{-\nu(n+a)}}{z^{(n+a)}} \times \\
  &\times\left(1+e^{-\nu}\right)^{\frac{1}{2}-c} \left(1-e^{-\nu}\right)^{c-a-b-\frac{1}{2}}
  \left[1+\mathcal{O}\left(\frac{1}{n}\right)\right] \nonumber 
\end{align}
with $\nu$ defined via the relation
\begin{equation} \label{nudefinition}
  e^{-\nu}=\frac{2-z-2(1-z)^{\frac{1}{2}}}{z} \, .
\end{equation}
We need to calculate
\begin{equation} 
  S[s,\theta,\omega]=\sum_{\lambda=2}^{\infty} \lambda \sin \left(\lambda s \right)
  \frac{\varphi_1^{(\lambda)}(\theta)}{{\varphi'}_1^{(\lambda)}(0)}
\end{equation}
with the ratio $\varphi_1^{(\lambda)}(\theta)/{\varphi'}_1^{(\lambda)}(0)$ given
in (\ref{RatioOf2Hyps}). Using the above expansion (\ref{GeneralAsymptoticOfHyp}) two times 
we find after numerous cancellations
\begin{align} \label{Sasymptotic}
  S[s,\theta,\omega]&\sim-\frac{4}{\omega}\left(\frac{1-z(\theta)}{1-z(0)}\right)^2
  \frac{\left[1+e^{-\nu(\theta)}\right]^{-\frac{1}{2}} \left[1-e^{-\nu(\theta)}\right]^{-\frac{5}{2}}}
  {\left[1+e^{-\nu(0)}\right]^{\frac{1}{2}} \left[1-e^{-\nu(0)}\right]^{-\frac{3}{2}}}
  \frac{z(0)^\frac{3}{2}}{z(\theta)^2}\frac{e^{-\frac{3\nu(\theta)}{2}}}{e^{-\frac{\nu(0)}{2}}} 
  \times \nonumber\\
  &\times \sum_{\lambda=2}^{\infty}\frac{\lambda \sin \left(\lambda s\right)}{\lambda-1}
  \left[\frac{e^{-\nu(\theta)}}{e^{-\nu(0)}}\right]^{\frac{\lambda}{2}} \, ,
\end{align}
where we used the following abbreviations
\begin{align}
  z(\theta)&=\tanh^2\left[\omega\left(\frac{\pi}{2}-\theta\right)\right] \, ,  \label{zdef}\\
  e^{-\nu(\theta)}&=\tanh^2\left[\frac{\omega}{2}\left(\frac{\pi}{2}-\theta\right)\right]
   \, .\label{nuresult}
\end{align}
Note that (\ref{nuresult}) is obtained after some lines of algebra by inserting (\ref{zdef}) in 
(\ref{nudefinition}). Still the sum in (\ref{Sasymptotic}) diverges for $\theta=0$ but it turns out that 
it can be performed analytically for $\theta>0$ so that after the summation the limit
$\theta \to 0$ exists. For the sake of a lighter notation we introduce the symbol $a$ by
\begin{equation} \label{adef}
  a\equiv\left[\frac{e^{-\nu(\theta)}}{e^{-\nu(0)}}\right]^{\frac{1}{2}}=
  \frac{\tanh\left[\frac{\omega}{2}\left(\frac{\pi}{2}-\theta\right)\right]}
       {\tanh\left(\frac{\omega\pi}{4}\right)} \, .
\end{equation}
It is now straightforward to evaluate the sum over $\lambda$:
\begin{align} 
  \sum_{\lambda=2}^{\infty}\frac{\lambda \sin \left(\lambda s\right)}{\lambda-1}
  a^\lambda&=\sum_{\lambda=2}^{\infty} \sin(\lambda s)\, a^\lambda+
  \sum_{\lambda=2}^{\infty}\frac{\sin \left(\lambda s\right)}{\lambda-1} a^\lambda \nonumber \\
  &=-a \sin s + \sum_{\lambda=1}^{\infty} \sin(\lambda s)\, a^\lambda+
  \sum_{\lambda=1}^{\infty}\frac{\sin \left[\left(\lambda+1\right) s\right]}{\lambda} a^{\lambda+1} \nonumber \\
  &=-a \sin s + \frac{1}{2}\frac{\sin s}{\frac{1}{2}\left(a+\frac{1}{a}\right)-\cos s}
  -\frac{a}{2} \sin s \, \ln \left( 1-2a\cos s+a^2\right) \nonumber \\
  &+ a\cos s \arctan\left[\frac{a \sin s}{1-a \cos s}\right] \, .
\end{align}
The full result can therefore be written as
\begin{align} \label{SasymptoticResult}
  S[s,\theta,\omega]&\sim-\frac{4}{\omega}\left(\frac{1-z(\theta)}{1-z(0)}\right)^2
  \frac{\left[1+e^{-\nu(\theta)}\right]^{-\frac{1}{2}} \left[1-e^{-\nu(\theta)}\right]^{-\frac{5}{2}}}
  {\left[1+e^{-\nu(0}\right]^{\frac{1}{2}} \left[1-e^{-\nu(0)}\right]^{-\frac{3}{2}}}
  \frac{z(0)^\frac{3}{2}}{z(\theta)^2}\frac{e^{-\frac{3\nu(\theta)}{2}}}{e^{-\frac{\nu(0)}{2}}} 
  \, a \sin s \, \times \\
  &\times  \left\{\frac{1}{2 a} \frac{1}{\frac{1}{2}\left(a+\frac{1}{a}\right)-\cos s}
   -\frac{1}{2} \ln \left( 1-2 a \cos s+a^2\right)+\cot s \, \arctan \left[\frac{a \sin s}{1-a \cos s}\right]
   -1 \right\} \, .\nonumber 
\end{align}
Although  the last result is given in closed form, we have to emphasize that its validity is restricted
to distances smaller than the size of the extra dimension $R$. 

It is now save to take the limit $\theta\to 0$ in (\ref{SasymptoticResult}). The result is 
\begin{align} \label{SasymptoticResultBrane}
  \lim_{\theta\to 0} S[s,\theta,\omega]&=-\frac{\sinh\left( \frac{\omega\pi}{2}\right)}{\omega}
   \sin s \left\{\frac{1}{4}\frac{1}{\sin^2(\frac{s}{2})}- \ln \left[2 \sin \left(\frac{s}{2}\right)\right]+
    \frac{\pi-s}{2} \cot s \; -1\right\} \, .
\end{align}
We note that it is the first term in the curly brackets on the right hand side of 
(\ref{SasymptoticResult}) which is responsible for reproducing the 
characteristic short distance singularity of the two-point function. The main result of 
this appendix is therefore:
\begin{align} 
  \lim_{\theta\to 0} S[s,\theta,\omega]&\sim-\frac{\sinh \left(\frac{\omega\pi}{2}\right)}{\omega}
  \left\{\frac{1}{s}+\frac{\pi}{2}+\mathcal{O}\left(s \ln s\right)\right\} \, .
\end{align}
The $5$-dimensional short distance singularity can even be obtained including the
extra dimension. Apart from the logarithmic term in the curly brackets of (\ref{SasymptoticResult}), 
all but the first one are finite at short distances (in $\theta$ and $s$). Developing
the denominator of this term, we recover the usual euclidean metric in $\mathbb{R}^4$:
\begin{equation} \label{5DNewton}
  \frac{1}{2}\left(a+\frac{1}{a}\right)-\cos s \sim 
  \frac{1}{2} \left[ \frac{\omega^2}{\sinh^2\left( \frac{\omega\pi}{2}\right)}\theta^2 +s^2 \right]+ \ldots .
\end{equation}
where the dots denote higher terms in $\theta$ and $s$.

\end{document}